\documentclass{aa}  

\usepackage[colorlinks=true, linkcolor = blue,
            urlcolor  = blue,
            citecolor = blue,
            anchorcolor = blue]{hyperref}
\usepackage{graphicx}
\usepackage{txfonts}
\usepackage{natbib}
\usepackage{enumitem} 
\usepackage{color}

\usepackage{cancel}
\bibpunct{(}{)}{;}{f}{}{,} % to follow the A&A style

\newcommand{\teff}{$T_{\mathrm{eff}}$}

\newcommand{\teffP}{$ T_{\text{eff}}^{\text{phot}} $}
\newcommand{\teffPCA}{$T_{\mathrm{eff}}^{PCA}$}
\newcommand{\tefflit}{$T_{\mathrm{eff}}^{lit}$}
\newcommand{\logg}{log \textit{g}}

\begin{document}

  \title{Faint solar analogs: at the limit of no reddening
  \thanks{Based on observations collected at Observat\'{o}rio do Pico dos Dias
  (OPD), operated by the Laborat\'orio Nacional de Astrof\'isica, CNPq,
  Brazil and on data from the ESO Science Archive Facility.}}

  \subtitle{Precise atmospheric parameters from moderate low resolution spectra}
    
   \author{R. E. Giribaldi\inst{1,2}
         \and
          G. F. Porto de Mello\inst{2}
         \and
          D. Lorenzo-Oliveira\inst{3}
         \and
          E. B. Am\^{o}res\inst{4}
         \and
          M. L. Ubaldo-Melo\inst{2}
          }

   \institute{ESO - European Southern Observatory, Karl-Schwarzchild-Strasse 2, 85748 Garching bei M\"{u}nchen, Germany \\
             \email{rescateg@eso.org, riano@astro.ufrj.br}
         \and
             Observat\'{o}rio do Valongo, Universidade Federal do Rio de Janeiro, Ladeira do Pedro Ant\^onio 43, 20080-090 Rio de Janeiro RJ, Brazil
         \and
             Universidade de S\~ao Paulo, Departamento de Astronomia do IAG/USP, Rua do Mat\~ao 1226, Cidade Universit\'aria, 05508-900 S\~ao Paulo SP, Brazil
         \and     
             UEFS, Departamento de F\'{i}sica, Av. Transnordestina, S/N, Novo Horizonte, Feira de Santana, CEP 44036-900, BA, Brazil.
             }
             
  % \date{Received September 15, 1996; accepted March 16, 1997}
    \date{Received  / Accepted }

% \abstract{}{}{}{}{} 
% 5 {} token are mandatory
 
  \abstract
   {The flux distribution of solar analogs is required for calculating the spectral albedo of Solar System bodies such as asteroids and trans-Neptunian objects.
   Ideally a solar analog should be comparably faint as the target of interest, but only few analogs fainter than $V\sim9$ were identified so far. 
   Only atmospheric parameters equal to solar guarantee a flux distribution equal to solar as well, while only photometric colors equal to solar do not. 
   Reddening is also a factor to consider when selecting faint analog candidates.}
   %when its effects equal the uncertainties of the photometric color measurements.}
   {We aim at implementing the methodology for identifying faint analogs at the limit of 
   precision allowed by current spectroscopic surveys.
   We quantify the precision attainable for the atmospheric parameters effective temperature (\teff), metallicity ([Fe/H]), surface gravity (\logg) when derived from moderate low resolution ($R=8000$) spectra with $S/N\sim100$.
   We estimate the significance of reddening at 100-300 pc from the Sun.}
   {We used the less precise photometry in the Hipparcos catalog to select potential analogs with $V\sim10.5$ (located at $\sim135$~pc).
   We calibrated \teff~and [Fe/H] as functions of equivalent widths of spectral indices 
   by means of the Principal Component Analysis (PCA) regression. We derive \logg, mass, radius, and age from the atmospheric parameters, Gaia parallaxes and evolutionary tracks. 
   We evaluated the presence of reddening for the candidates by underestimations of photometric \teff~with respect to those derived by spectral indices. These determinations are validated with extinction maps.}
   {We obtained atmospheric parameters \mbox{\teff/[Fe/H]/\logg} with precision of \mbox{97 K/0.06 dex/0.05 dex}. 
   From 21 candidates analysed, we identify five solar analogs:
   HIP~991, HIP~5811, and HIP~69477
   have solar parameters within 1$\sigma$ errors, and HIP~55619 and HIP~61835 within 2$\sigma$ errors. 
   Other six stars 
   %HIP~6089, HIP~18941, HIP~31845, HIP~48272, HIP~56870 and HIP~107605, 
   have \teff~close to solar but slightly lower [Fe/H]. 
   Our analogs show no evidence of reddening  but for four stars, which present $E(B-V) \geq 0.06$~mag, translating to at least a 200~K decrease in photometric \teff.}
   {}


   \maketitle
%
%________________________________________________________________

\section{Introduction}

The Sun remains the primary and most fundamental reference object in stellar astrophysics, being the golden standard for a variety of physical and chemical properties and still the sole star for which we access both extensively and accurately important fundamental parameters \citep{porto2014,ram2009,mel2006,cayrel1996}. 
On the one hand, the search for stars identical to the Sun in their physical properties, the so called solar twins, did provide an interesting contextualization of the properties of the ``Sun as a star''. For example, concerning its age, chromospheric activity, and detailed chemical abundance \citep{Mel2014,li2012,nasc2009,porto1997} among other quantities. 
On the other hand, a very relevant
motivation to find and characterize stars that reproduce the solar spectrophotometric properties, something that solar twins are naturally expected to do, is the need to know reliable reference stars, observable at night under the same conditions as other targets of interest \citep{soubiran2004,porto2014}. 
Hence the need to look for solar analogs, stars that closely reproduce the solar flux distribution, and may thus act as solar surrogates in the night sky.

According to the traditional definition of \citet{cayrel1996}, 
solar analogs are solar-type stars with atmospheric parameters effective temperature (\teff), metallicity ([Fe/H]\footnote{[A/B] = $ \log{\left( \frac{N(\text{A})}{N(\text{B})} \right )_\text{Star}} - \log{\left( \frac{N(\text{A})}{N(\text{B})} \right )_\text{Sun}} $, where $ N $ denotes the
     number abundance of a given element.}), and surface gravity (\logg)
similar to those of the Sun within specified uncertainty criteria, and therefore they present a solar flux distribution.
Stars with photometric colors equal to solar are sometimes called solar analogs in the literature, but we remark 
that the use of this working definition should be taken with care because solar photometric colors, 
only, hardly imply a solar flux distribution.
This is the reason why their atmospheric parameters must be proven to be solar by spectroscopic techniques.
Solar analogs may serve as calibrating objects when the solar flux distribution needs to be observed at night, that is, as ``solar proxies'' or ``solar surrogates''. 
Ideally they should be
known to magnitudes comparably faint to the targets of interest in order to record the instrumental signature in the spectra of both the target and the proxy.
Furthermore, the availability of a list of solar analogs well spread in the sky allows the users to choose a solar proxy close to the target, to be observed with a similar airmass to record the same telluric features as in the target's spectrum.
A proper solar proxy then guarantees  the complete removal of the solar signature, of the instrumental signature, and of the telluric features, which is essential for recovering accurately the body's albedo, whose shape and inclination are used for taxonomy \citep[e.g.,][]{chapman1975,tholen1989}.
Since Solar System bodies such as trans-Neptunian objects with $V \sim 15$, or fainter, are routinely observed nowadays, 
it is reasonable to require solar proxies with, for example $V =$ 13-14. Such proxies are at least 10 times brighter than common targets but should still allow convenient corrections.
 
\citet{porto2014} recently provided a sizable list of
solar analogs, characterized both photometrically and
spectroscopically, and well distributed in the night sky, widely
extending both in quality and quantity the initial work of
\citet{hardorp1982}, who
provided a first impetus on the search for solar analogs.
Surprisingly, Hardorp's lists are still being referred to nowadays.
However, the lists of Porto de Mello et al. reach no
fainter than $V\sim9$, not much better than Hardorp's, only
sampling stars within 50~pc of the Sun. 
This magnitude range is too bright for telescopes of the 8-10m class. 
An example that illustrates the need for fainter solar analogs is the use of the stars BD+00~3383 ($V=10.50$) and HD~11532 ($V=9.71$). 
They both showed acceptable performances as solar proxies although no detailed spectroscopy was applied to them for determining their atmospheric parameters.
They were used for recovering albedos from the infrared to the visible \citep[e.g.,][]{merlin2017, dumas2011, alvarez_candal2008}, and the UV \citep[e.g.,][]{snodgrass2017}. 
For a solar-type star, the flux variation as a function of the atmospheric parameters from the infrared to the visible keeps nearly a constant shape, but this no longer applies from $\lambda$5000 downwards to the UV, a region much more sensitive to \teff, [Fe/H], and \logg$\,$ shifts, in this order. 
For example, Fig~1 in \citet{fernley1996} shows that
a variation of $-200$~K from the solar \teff~increases the flux by $\sim15$\%
at 4000~\AA$\,$with respect to that at 5000~\AA.
In cases like this, solar analogs with atmospheric parameters precisely close to solar are advisable in order to assert minimum influence on the intrinsic shape of the target's albedo.

In the present work we implement methods to identify faint solar analogs. 
The definition of ``faint'' is subjective because it has to conform to the requirements of the users, 
or to the faintest analogs identified so far. 
For example, stars that were considered faint in the Hipparcos catalog are definitely 
not so for present Gaia standards; 20 years of technological advances allow much deeper sky prospecting.
Since we hunt solar analogs, the definition of faint we adopt conforms to the 
apparent magnitude of the stars whose photometric, astrometric, and spectroscopic available 
data have the minimum quality to determine their atmospheric parameters with reasonable precision and present techniques:
100~K/0.05~dex/0.05~dex in \teff/[Fe/H]/\logg.
We employ as our initial screening photometric data from Hipparcos, obviously not current, though they were so at the time our survey started. Much more precise photometric and astrometric data were made available by Gaia \citep{Gai(b)}.
However, the stars with the less precise photometry in Hipparcos, close to this catalog's completeness limit, those with $V\sim10.5$, are still competitive as reasonably faint solar analogs. The methods here implemented can be readily applied to spectra with similar characteristics acquired by telescopes of 8-10m, corresponding to stars of $V = 16$-18.

Interstellar extinction arises as an additional problem for the selection of candidates as they become increasingly fainter and farther away. 
Solar analogs are most probably located in the Galactic thin disk because the metallicity distribution 
of this population is essentially solar \citep[e.g.,][]{adi2013}.
The scale height of the thin disk is estimated at $\sim300$~pc
\citep[][and references therein]{juric2008},
thus at longer distances, a more productive search would be performed by pointing to the Galactic plane than to the poles. 
At the same time, pointing to the plane implies candidates with more attenuated magnitudes and more reddened colors,
thus precise corrections for Galactic layers must be applied.
Such solar analogs will probably not satisfy the need for solar surrogates, either photometrically or spectroscopically, from the blue limit of the $H$ band to shorter wavelengths because extinction increases fast from there \citep[e.g.,][Fig.~10]{gordon2003}. 

Some faint solar analogs and twins were already identified, for example Inti~1 with $V = 12.86$ \citep{galarza2016},
KIC 10971974 with $V = 11.05$ \citep{beck2017}, and those in the M67 cluster with $V \sim 14.60$ \citep{Pasquini2008, Onehag2011}.
Here we provide a short list of solar analogs as products of testing our methods.
They should be subsequently submitted to more precise spectroscopic analyses to better determine 
their fundamental parameters derive other quantities as rotation, detailed chemical composition, 
magnetic activity, and asteroseismological properties.

This paper is organized as follows: In Sect.~\ref{sec2} we describe the selection criterion of the candidates. 
In Sect.~\ref{sec:reduction} we describe the data reduction.
Sect.~\ref{sec:indices} describes the Principal Component Analisis (PCA) regression applied to spectral indices.
Sect.~\ref{sec:parameters} presents the fundamental parameters derived for the candidates.
In Sect.~\ref{sect:redd} we determine the influence of reddening on photometric colors.
In Sect.~\ref{best} we summarize the information obtained for the best solar analogs identified, and finally in Sect.~\ref{sec:conclusions}, we synthesize our conclusions.

%__________________________________________________________________
%SEC: SELECTING SAMPLE OF FAINT ANALOG CANDIDATES
%__________________________________________________________________
\section{Selecting the sample of faint analog candidates}
\label{sec2}

This survey was launched by the time the Hipparcos catalog \citep{Perryman} was the reference for the most precise parallaxes, colors, and 
magnitudes for solar-type stars, and the procedure we employed consider this fact.

We start our search by selecting candidates by colors (proxies of \teff~and [Fe/H]) and absolute magnitudes (proxy of \logg), 
the observable quantities that allow a gross selection. 
The colors of widespread
use, and having available several \teff~calibrations,  
are $B-V$ and ($B-V$)$^{\rm Ty}$
from the Johnson and Tycho \citep{Hoeg} systems.
The initial procedure follows closely the one in \citet{porto2014}: ``boxes'' were prospected around the solar colors and absolute magnitudes ($B-V$)$_\odot = 0.654$, ($B-V$)$_{\odot}^{\rm
Ty} = 0.733$, M$_{\mathrm{V}_\odot} = 4.82$, \mbox{M$_{\rm
V_{\odot}}^{\rm Ty} = 4.88$}.
Hipparcos is complete up to $V \sim 9$, but still lists fainter stars in decreasing
degrees of completeness down to $V \sim 11$. 
We chose Hipparcos for the sample selection because it has more precise parallaxes than Tycho, which permits a more reliable selection based in magnitudes, although Tycho goes deeper, being complete down to $V^{Ty} = 10$, and still 90\% complete down to $V^{Ty} \sim 10.5$.
 
The final list of candidates to be analyzed
spectroscopically should have a size compatible with the
accomplishment of this project in the Observat\'orio Pico dos Dias
(OPD) operated by Laboratório Nacional de Astrofisica (LNA/CNPq) 
within a period of a few years. These practical order
considerations constrain the candidate list to, at most, some tens
of objects.
We emphasize, however, that while we used Hipparcos for the sample selection, the determination of \logg, mass, radius, age, and reddening were updated
with the parallaxes of Gaia DR2
\citep{Gai(b),Gai(c)}.

We started with some coarse tests to gauge the size of the sample.
Solar-type stars with $V$ between
10.5 and 11.2 were considered, the faint limit of Hipparcos. 
The dimensions of the boxes around the solar colors and absolute magnitudes were set by the mean of the 1$\sigma$ errors of all stars 
contained within the box, self-consistently.
We worked simultaneously with boxes around the Johnson and Tycho solar colors and absolute magnitudes, and we kept stars within 2$\sigma$ of the boxes' centers. The average uncertainties of color and absolute magnitude for the
stars contained in the box, respectively, are very similar to the
uncertainty values used to define the boxes in the first place, representative uncertainties being:
       \[
      \begin{array}{lp{0.5\linewidth}}
         <\sigma(B-V)>  & = 0.07     \\
         <\sigma(B-V)^{Ty}>  & = 0.12      \\
         <\sigma(M_{V})> & = 0.80   \\
         <\sigma(M_{V})^{Ty}> & = 0.80    \\
      \end{array}
      \]
These tests constrained samples with around 300 stars, which we decreased by
considering only those objects, within these initial 2$\sigma$
boxes defined by average errors, for which the 
individual uncertainty implied in a 2$\sigma$ agreement with the
solar values defining the centers of the boxes. This second sample totalled 203 stars, average errors being:
      \[
      \begin{array}{lp{0.5\linewidth}}
         <\sigma(B-V)>  & = 0.067     \\
         <\sigma(B-V)^{Ty}>  & = 0.112      \\
         <\sigma(M_{V})> & = 0.609   \\
         <\sigma(M_{V})^{Ty}> & = 0.682    \\
      \end{array}
      \]
Finally, we fine tuned this subsample by retaining only those
objects for which the individual errors were no larger than the
average errors defined for each box, thus a 1$\sigma$
criterion, applying the cuts stepwise in the M$_{\rm V_{\odot}}$,
M$_{\rm V_{\odot}}^{\rm Ty}$, ($B-V$)$_\odot^{\rm Ty}$ and
($B-V$)$_\odot$ dimensions, in this order. 
We have purposefully disregarded reddening in the 
selection process in order to gauge its influence in the method of
selection.

\begin{figure*}
      \centering
      \includegraphics[width=8.5cm, angle = 90]{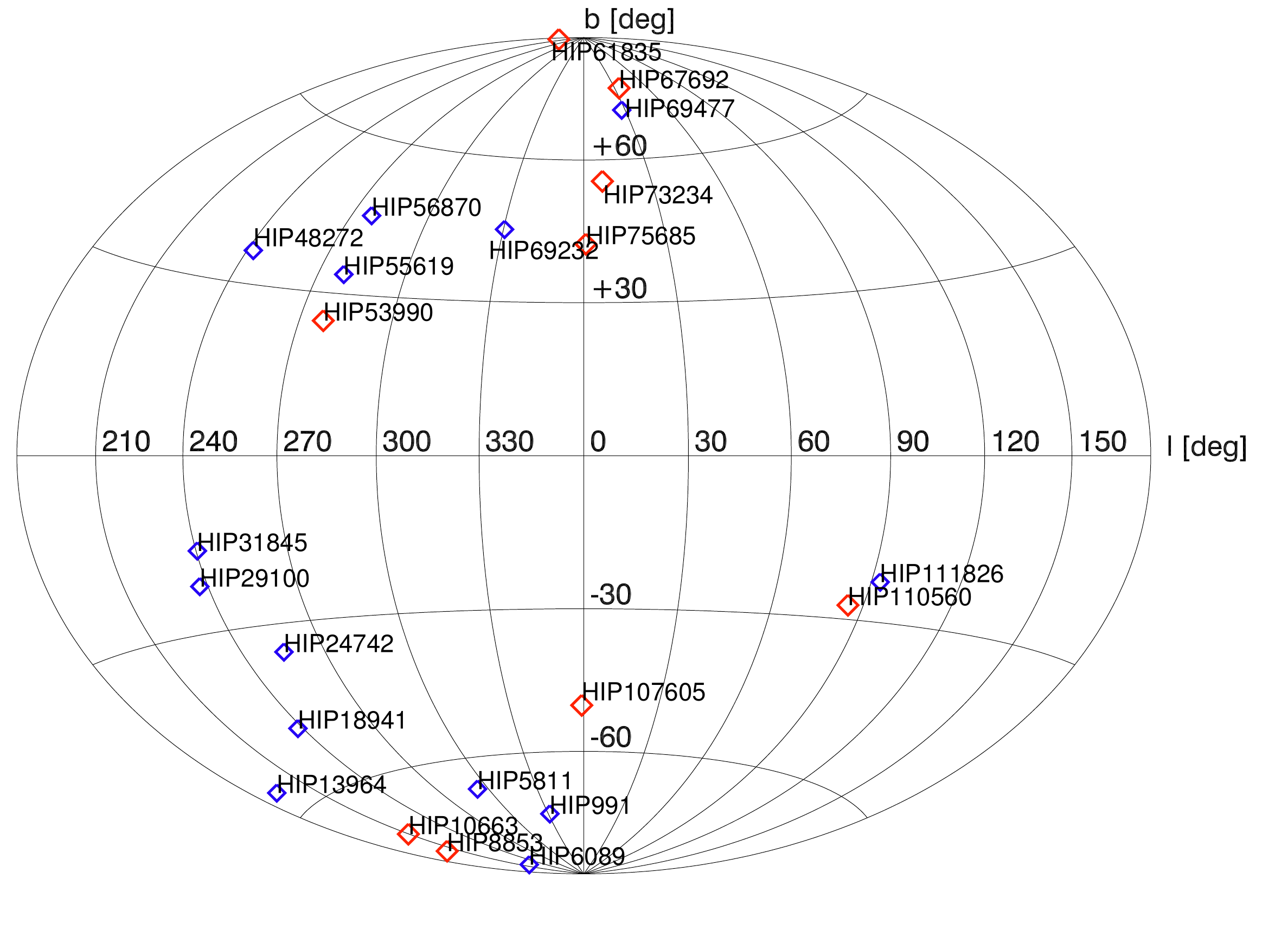}
      \caption{Distribution of candidates in Galactic coordinates. 
      Stars with distances $ < $ 150 pc are the blue diamonds, and with distances
      $ > $ 150 pc, red diamonds.}
      \label{Fig0}
      \end{figure*}

The selected candidate sample contains 41 stars, and it is listed in Table~\ref{tab:canddtstars}.
It displays the stellar photometric and astrometric measurements as shown in the catalogs Hipparcos, Two Micron All Sky Survey \citep[][]{2MASS},
and Gaia DR2. 
The table is divided in two parts, the first one lists the observed stars, 
that we refer henceforth simply as ``candidates'', for which the $S/N$ is noted.
We also show in Fig.~\ref{Fig0} the spatial distribution of the candidates in galactic coordinates. 
No candidates are located towards the galactic plane, thus their reddening is expected to be low,
with some exceptions as found in Sect.~\ref{sect:redd}.

     \begin{table*}
     \centering
     \scriptsize 
     \caption{Photometric and astrometric data of the candidates. Stars for which no S/N is given in column 11 were photometrically selected but not spectroscopically observed.
     The first column lists the Hipparcos number. Columns 2 and 3 display the coordinates, right ascension and declination.
     Columns 4 and 5 list $ B-V $ in Johnson and Tycho systems.
     Columns 6 to 9 list the magnitudes in the indicated photometric bands. 
     $ V $ was extracted from the Hipparcos catalog to which we associated the error of the same band in the
     Tycho system. 
     Column 10 lists parallaxes from Gaia DR2 \protect{\citep{Gai(c)}.}
     Column 11 lists the $S/N$ of the acquired spectra.}
     \label{tab:canddtstars}
     \begin{tabular}{c c c c c c c c c c c}
     \hline\hline 
     HIP & $ RA $ & $ DEC $ & $ (B-V) $ & $ (B-V)^{Ty} $ & $ V $ & $ J $ & $ H $ & $ K_s $ & \textit{parallax} (mas) & $S/N$ \\
     %	&	&	 &	 &	&   &  &  &   & (pc) &  \\
     \hline 
     991	&	 $ 00:12:18 $ 	&	 $ -40:38:44 $ 	&	 $ 0.600 \pm 0.061 $ 	&	 $ 0.647 \pm 0.070 $ 	&	 $10.58 \pm 0.047$ 	&	 $ 9.476 \pm 0.026 $ 	&	 $ 9.136 \pm 0.026 $ 	&	 $ 9.067 \pm 0.024 $ 	&	$ 7.0441 \pm 0.0326 $	&	 108\\
     5811	&	 $ 01:14:33 $ 	&	 $ -49:54:12 $ 	&	 $ 0.700 \pm 0.004 $ 	&	 $ 0.767 \pm 0.090 $ 	&	 $10.62 \pm 0.055$ 	&	 $ 9.458 \pm 0.032 $ 	&	 $ 9.106 \pm 0.033 $ 	&	 $ 9.077 \pm 0.033 $ 	&	$ 7.5506 \pm 0.0271 $	&	 112\\
     6089	&	 $ 01:18:11 $ 	&	 $ -27:36:17 $ 	&	 $ 0.661 \pm 0.015 $ 	&	 $ 0.647 \pm 0.094 $ 	&	 $10.55 \pm 0.061$ 	&	 $ 9.353 \pm 0.030 $ 	&	 $ 8.990 \pm 0.076 $ 	&	 $ 8.912 \pm 0.019 $ 	&	$ 8.7481 \pm 0.0412 $	&	 126\\
     8853	&	 $ 01:53:51 $ 	&	 $ -23:29:52 $ 	&	 $ 0.530 \pm 0.020 $ 	&	 $ 0.563 \pm 0.083 $ 	&	 $10.63 \pm 0.058$ 	&	 $ 9.672 \pm 0.024 $ 	&	 $ 9.408 \pm 0.022 $ 	&	 $ 9.404 \pm 0.023 $ 	&	$ 5.2837 \pm 0.0386 $	&	 90\\
     10663	&	 $ 02:17:13 $ 	&	 $ -24:23:56 $ 	&	 $ 0.570 \pm 0.020 $ 	&	 $ 0.522 \pm 0.100 $ 	&	 $10.62 \pm 0.072$ 	&	 $ 9.615 \pm 0.023 $ 	&	 $ 9.379 \pm 0.022 $ 	&	 $ 9.289 \pm 0.021 $ 	&	$ 3.6740 \pm 0.0386 $	&	 121\\
     13964	&	 $ 02:59:49 $ 	&	 $ -11:20:42 $ 	&	 $ 0.556 \pm 0.015 $ 	&	 $ 0.548 \pm 0.093 $ 	&	 $10.53 \pm 0.065$ 	&	 $ 9.457 \pm 0.023 $ 	&	 $ 9.058 \pm 0.022 $ 	&	 $ 8.967 \pm 0.020 $ 	&	$ 9.0105 \pm 0.0560 $	&	 210\\
     18941	&	 $ 04:03:36 $ 	&	 $ -36:10:40 $ 	&	 $ 0.590 \pm 0.020 $ 	&	 $ 0.564 \pm 0.078 $ 	&	 $10.52 \pm 0.055$ 	&	 $ 9.376 \pm 0.027 $ 	&	 $ 9.093 \pm 0.024 $ 	&	 $ 9.005 \pm 0.021 $ 	&	$ 7.0380 \pm 0.0221 $	&	 114\\
     24742	&	 $ 05:18:19 $ 	&	 $ -48:52:12 $ 	&	 $ 0.529 \pm 0.032 $ 	&	 $ 0.518 \pm 0.091 $ 	&	 $10.67 \pm 0.063$ 	&	 $ 9.514 \pm 0.029 $ 	&	 $ 9.167 \pm 0.022 $ 	&	 $ 9.095 \pm 0.021 $ 	&	$ 7.1532 \pm 0.0184 $	&	 103\\
     29100$^{*}$	&	 $ 06:08:17 $ 	&	 $ -30:40:05 $ 	&	 $ 0.611 \pm 0.003 $ 	&	 $ 0.657 \pm 0.081 $ 	&	 $10.56 \pm 0.053$ 	&	 $ 9.429 \pm 0.022 $ 	&	 $ 9.092 \pm 0.022 $ 	&	 $ 9.054 \pm 0.019 $ 	&	$ 7.8400 \pm 0.2100 $	&	 134\\
     31845	&	 $ 06:39:30 $ 	&	 $ -31:25:50 $ 	&	 $ 0.626 \pm 0.015 $ 	&	 $ 0.777 \pm 0.086 $ 	&	 $10.51 \pm 0.052$ 	&	 $ 9.191 \pm 0.023 $ 	&	 $ 8.832 \pm 0.044 $ 	&	 $ 8.789 \pm 0.024 $ 	&	$ 9.0866 \pm 0.0241 $	&	 112\\
     48272	&	 $ 09:50:29 $ 	&	 $ -04:57:37 $ 	&	 $ 0.536 \pm 0.003 $ 	&	 $ 0.595 \pm 0.107 $ 	&	 $10.51 \pm 0.072$ 	&	 $ 9.387 \pm 0.023 $ 	&	 $ 9.096 \pm 0.023 $ 	&	 $ 8.997 \pm 0.020 $ 	&	$ 7.3291 \pm 0.0359 $	&	 92\\
     55619	&	 $ 11:23:43 $ 	&	 $ -25:06:30 $ 	&	 $ 0.667 \pm 0.004 $ 	&	 $ 0.762 \pm 0.092 $ 	&	 $10.55 \pm 0.058$ 	&	 $ 9.344 \pm 0.027 $ 	&	 $ 8.937 \pm 0.026 $ 	&	 $ 8.884 \pm 0.021 $ 	&	$ 7.6182 \pm 0.0409 $	&	 121\\
     56870	&	 $ 11:39:34 $ 	&	 $ -14:04:34 $ 	&	 $ 0.645 \pm 0.003 $ 	&	 $ 0.872 \pm 0.095 $ 	&	 $10.53 \pm 0.055$ 	&	 $ 9.260 \pm 0.024 $ 	&	 $ 8.931 \pm 0.023 $ 	&	 $ 8.839 \pm 0.024 $ 	&	$ 9.2905 \pm 0.0356 $	&	 122\\
     61835	&	 $ 12:40:17 $ 	&	 $ +27:46:34 $ 	&	 $ 0.588 \pm 0.015 $ 	&	 $ 0.527 \pm 0.103 $ 	&	 $10.80 \pm 0.073$ 	&	 $ 9.719 \pm 0.023 $ 	&	 $ 9.427 \pm 0.027 $ 	&	 $ 9.373 \pm 0.022 $ 	&	$ 5.5279 \pm 0.0527 $	&	 143\\
     67692	&	 $ 13:51:59 $ 	&	 $ +26:38:11 $ 	&	 $ 0.750 \pm 0.015 $ 	&	 $ 0.906 \pm 0.102 $ 	&	 $10.94 \pm 0.060$ 	&	 $ 9.587 \pm 0.022 $ 	&	 $ 9.204 \pm 0.019 $ 	&	 $ 9.139 \pm 0.022 $ 	&	$ 3.4312 \pm 0.0612 $	&	 92\\
     69232	&	 $ 14:10:27 $ 	&	 $ -13:56:04 $ 	&	 $ 0.605 \pm 0.025 $ 	&	 $ 0.647 \pm 0.107 $ 	&	 $10.67 \pm 0.071$ 	&	 $ 9.404 \pm 0.023 $ 	&	 $ 9.055 \pm 0.022 $ 	&	 $ 8.961 \pm 0.024 $ 	&	$ 7.1930 \pm 0.0500 $	&	 81\\
     69477	&	 $ 14:13:25 $ 	&	 $ +23:54:03 $ 	&	 $ 0.562 \pm 0.066 $ 	&	 $ 0.603 \pm 0.075 $ 	&	 $10.53 \pm 0.052$ 	&	 $ 9.307 \pm 0.019 $ 	&	 $ 9.010 \pm 0.021 $ 	&	 $ 8.960 \pm 0.024 $ 	&	$ 8.2887 \pm 0.0321 $	&	 114\\
     73234	&	 $ 14:58:03 $ 	&	 $ +09:24:03 $ 	&	 $ 0.680 \pm 0.061 $ 	&	 $ 0.743 \pm 0.077 $ 	&	 $10.59 \pm 0.050$ 	&	 $ 9.448 \pm 0.023 $ 	&	 $ 9.143 \pm 0.023 $ 	&	 $ 9.078 \pm 0.023 $ 	&	$ 5.3550 \pm 0.0761 $	&	 87\\
     75685	&	 $ 15:27:42 $ 	&	 $ -02:45:18 $ 	&	 $ 0.730 \pm 0.015 $ 	&	 $ 0.872 \pm 0.099 $ 	&	 $10.51 \pm 0.060$ 	&	 $ 9.186 \pm 0.024 $ 	&	 $ 8.870 \pm 0.042 $ 	&	 $ 8.810 \pm 0.024 $ 	&	$ 6.3801 \pm 0.0337 $	&	 72\\
     107605	&	 $ 21:47:41 $ 	&	 $ -41:51:17 $ 	&	 $ 0.640 \pm 0.020 $ 	&	 $ 0.664 \pm 0.090 $ 	&	 $10.60 \pm 0.060$ 	&	 $ 9.498 \pm 0.022 $ 	&	 $ 9.243 \pm 0.027 $ 	&	 $ 9.180 \pm 0.021 $ 	&	$ 6.4888 \pm 0.0456 $	&	 170\\
     111826	&	 $ 22:39:01 $ 	&	 $ +32:18:03 $ 	&	 $ 0.762 \pm 0.065 $ 	&	 $ 0.850 \pm 0.086 $ 	&	 $10.53 \pm 0.053$ 	&	 $ 9.216 \pm 0.027 $ 	&	 $ 8.817 \pm 0.026 $ 	&	 $ 8.796 \pm 8.786 $ 	&	$ 8.4807 \pm 0.0326 $	&	 125\\
     \hline
13052	&	 $02:47:45$ 	&	 $+80:15:54$ 	&	 0.784 $\pm$ 0.062 	&	 0.899 $\pm$ 0.080 	&	 10.53 $\pm$ 0.047 	&	 9.155 $\pm$ 0.022 	&	 8.811 $\pm$ 0.029 	&	 8.711 $\pm$ 0.025	&	11.1187 $\pm$ 0.0436	&	---	\\
16294	&	 $03:30:03$ 	&	 $+51:30:43$ 	&	 0.520 $\pm$ 0.020 	&	 0.729 $\pm$ 0.086 	&	 10.56 $\pm$ 0.054 	&	 9.337 $\pm$ 0.020 	&	 9.119 $\pm$ 0.026 	&	 9.053 $\pm$ 0.020	&	5.4415 $\pm$ 0.0526	&	---	\\
17514	&	 $03:45:00$ 	&	 $-38:51:33$ 	&	 0.598 $\pm$ 0.015 	&	 0.782 $\pm$ 0.010 	&	 10.64 $\pm$ 0.063 	&	 9.447 $\pm$ 0.021 	&	 9.111 $\pm$ 0.024 	&	 9.017 $\pm$ 0.023	&	7.6114 $\pm$ 0.0324	&	---	\\
46072	&	 $09:23:41$ 	&	 $+65:48:31$ 	&	 0.675 $\pm$ 0.044 	&	 0.737 $\pm$ 0.055 	&	 10.53 $\pm$ 0.035 	&	 9.329 $\pm$ 0.020 	&	 9.062 $\pm$ 0.017 	&	 8.979 $\pm$ 0.016	&	7.0718 $\pm$ 0.0272	&	---	\\
53442	&	 $10:55:58$ 	&	 $+29:19:13$ 	&	 0.552 $\pm$ 0.067 	&	 0.592 $\pm$ 0.076 	&	 10.51 $\pm$ 0.051 	&	 9.362 $\pm$ 0.021 	&	 9.041 $\pm$ 0.016 	&	 8.955 $\pm$ 0.018	&	7.5127 $\pm$ 0.0845	&	---	\\
53990	&	 $11:02:39$ 	&	 $-32:44:17$ 	&	 0.550 $\pm$ 0.020 	&	 0.732 $\pm$ 0.090 	&	 10.67 $\pm$ 0.057 	&	 9.553 $\pm$ 0.024 	&	 9.336 $\pm$ 0.024 	&	 9.229 $\pm$ 0.021	&	5.3337 $\pm$ 0.0431	&	---	\\
55229	&	 $11:18:36$ 	&	 $+50:44:55$ 	&	 0.688 $\pm$ 0.062 	&	 0.753 $\pm$ 0.078 	&	 10.76 $\pm$ 0.048 	&	 9.580 $\pm$ 0.022 	&	 9.269 $\pm$ 0.028 	&	 9.235 $\pm$ 0.023	&	4.4272 $\pm$ 0.0394	&	---	\\
55809	&	 $11:26:11$ 	&	 $+53:32:39$ 	&	 0.654 $\pm$ 0.049 	&	 0.729 $\pm$ 0.065 	&	 10.50 $\pm$ 0.042 	&	 9.257 $\pm$ 0.019 	&	 9.001 $\pm$ 0.031 	&	 8.852 $\pm$ 0.022	&	5.3923 $\pm$ 0.0330	&	---	\\
59223	&	 $12:08:47$ 	&	 $+30:56:33$ 	&	 0.542 $\pm$ 0.065 	&	 0.580 $\pm$ 0.074 	&	 10.51 $\pm$ 0.051 	&	 9.464 $\pm$ 0.022 	&	 9.220 $\pm$ 0.021 	&	 9.162 $\pm$ 0.017	&	6.2389 $\pm$ 0.0398	&	---	\\
59369	&	 $12:10:49$ 	&	 $+32:44:54$ 	&	 0.573 $\pm$ 0.067 	&	 0.615 $\pm$ 0.076 	&	 10.58 $\pm$ 0.051 	&	 9.523 $\pm$ 0.027 	&	 9.174 $\pm$ 0.028 	&	 9.191 $\pm$ 0.019	&	5.5150 $\pm$ 0.0364	&	---	\\
60523	&	 $12:24:25$ 	&	 $+53:26:54$ 	&	 0.680 $\pm$ 0.055 	&	 0.743 $\pm$ 0.069 	&	 10.77 $\pm$ 0.043 	&	 9.650 $\pm$ 0.027 	&	 9.356 $\pm$ 0.026 	&	 9.261 $\pm$ 0.018	&	6.3396 $\pm$ 0.0294	&	---	\\
61957	&	 $12:41:51$ 	&	 $+26:49:47$ 	&	 0.585 $\pm$ 0.015 	&	 0.511 $\pm$ 0.075 	&	 10.54 $\pm$ 0.052 	&	 9.540 $\pm$ 0.023 	&	 9.264 $\pm$ 0.021 	&	 9.220 $\pm$ 0.020	&	4.4171 $\pm$ 0.1482	&	---	\\
63588	&	 $13:01:51$ 	&	 $+27:20:15$ 	&	 0.594 $\pm$ 0.015 	&	 0.802 $\pm$ 0.112 	&	 10.70 $\pm$ 0.069 	&	 9.525 $\pm$ 0.026 	&	 9.271 $\pm$ 0.034 	&	 9.131 $\pm$ 0.020	&	6.0707 $\pm$ 0.0460	&	---	\\
67215	&	 $13:46:26$ 	&	 $+82:31:46$ 	&	 0.695 $\pm$ 0.065 	&	 0.783 $\pm$ 0.086 	&	 10.52 $\pm$ 0.054 	&	 9.474 $\pm$ 0.020 	&	 9.222 $\pm$ 0.017 	&	 9.157 $\pm$ 0.017	&	6.2334 $\pm$ 0.0259	&	---	\\
69554	&	 $14:14:14$ 	&	 $+38:19:58$ 	&	 0.723 $\pm$ 0.066 	&	 0.819 $\pm$ 0.087 	&	 10.79 $\pm$ 0.053 	&	 9.568 $\pm$ 0.020 	&	 9.256 $\pm$ 0.016 	&	 9.209 $\pm$ 0.016	&	7.2600 $\pm$ 0.0256	&	---	\\
73854	&	 $15:05:37$ 	&	 $+45:23:49$ 	&	 0.724 $\pm$ 0.064 	&	 0.800 $\pm$ 0.083 	&	 10.53 $\pm$ 0.052 	&	 9.435 $\pm$ 0.021 	&	 9.164 $\pm$ 0.019 	&	 9.098 $\pm$ 0.020	&	7.3037 $\pm$ 0.0241	&	---	\\
74061	&	 $15:08:09$ 	&	 $+39:58:12$ 	&	 0.633 $\pm$ 0.064 	&	 0.700 $\pm$ 0.085 	&	 10.58 $\pm$ 0.055 	&	 9.462 $\pm$ 0.021 	&	 9.098 $\pm$ 0.021 	&	 9.008 $\pm$ 0.014	&	5.3679 $\pm$ 0.0723	&	---	\\
76272	&	 $15:34:45$ 	&	 $+62:16:44$ 	&	 0.592 $\pm$ 0.065 	&	 0.637 $\pm$ 0.075 	&	 10.52 $\pm$ 0.051 	&	 9.684 $\pm$ 0.021 	&	 9.315 $\pm$ 0.017 	&	 9.216 $\pm$ 0.020	&	6.2103 $\pm$ 0.0269	&	---	\\
102416	&	 $20:45:13$ 	&	 $+60:19:35$ 	&	 0.642 $\pm$ 0.064 	&	 0.712 $\pm$ 0.085 	&	 10.52 $\pm$ 0.055 	&	 9.341 $\pm$ 0.023 	&	 9.027 $\pm$ 0.031 	&	 8.970 $\pm$ 0.019	&	7.8116 $\pm$ 0.0293	&	---	\\
110560	&	 $22:23:49$ 	&	 $+24:23:34$ 	&	 0.573 $\pm$ 0.016 	&	 0.773 $\pm$ 0.097 	&	 10.64  $\pm$ 0.059 	&	 9.440$\pm$  0.022 	&	 9.172 $\pm$ 0.021 	&	 9.106 $\pm$ 0.018	&	5.0932 $\pm$ 0.0383	&	---	\\
\hline
\multicolumn{11}{l}{*The parallax of this candidate was extracted from the Gaia DR1 catalog \protect{\citep{Gai(b),Gai(a)}}.}\\
     \end{tabular}
     \end{table*}  

%__________________________________________________________________
%SEC: OBSERVATIONS AND DATA REDUCTION
%__________________________________________________________________
\section{Observations and data reduction}
\label{sec:reduction}
   Spectroscopic observations were performed with the long-slit coudé spectrograph, coupled
   to the 1.60m telescope of OPD in  six missions from 1998 to 2013. 
   The spectra cover a range of 500 \AA\ centered in $ \lambda $6563 (H$ \alpha $), 
   and have a nominal resolution of $R = \lambda/\Delta \lambda \sim $ 8000. 
   The signal-to-noise ratio ($S/N$) of the spectra spans between 70 and 220 for the candidates, and between 70 and 810 for the
   calibration stars, see Tables~\ref{tab:canddtstars} and \ref{tab:cab_stars}, respectively.
   
   The data reduction was carried out by the standard procedure using 
   IRAF\footnote{\textit{Image Reduction and Analysis Facility} (IRAF) is distributed by 
   the National Optical Astronomical Observatories (NOAO), which is operated by the
   Association of Universities for Research in Astronomy (AURA), Inc., under contract to 
   the National Science Foundation (NSF).}, i.e. 
   for one-dimensional spectra extraction, bias and flat-field corrections were performed prior to 
   background and scattered light subtraction.
   The pixel-to-wavelength calibration was obtained by comparing the spectra 
   of stars with a Thorium-Argon lamp spectra acquired in the same night of the observations. 
   Doppler corrections were applied for all spectra and  
   continuum normalizations were performed by fitting low-order polynomials to the highest flux 
   regions following a systematic procedure.
   
%__________________________________________________________________
%SEC: ATMOSPHERIC PARAMETERS
%__________________________________________________________________
\section{Calibration of spectral indices}
\label{sec:indices}
     In order to determine the atmospheric parameters of the candidates, 
     we built a calibration by means of 
     %a Principal Component Regression (PCR) 
     the Principal Component Analysis (PCA) applied to the equivalent width (EW) of the spectral indices. At $R = 8000$, 
     individual metallic lines are not resolved, thus, the determination of atmospheric 
     parameters using spectroscopic techniques such as the excitation and ionization equilibrium of Fe lines, and 
     Balmer-lines fitting is not possible. Alternatively, spectral indices have been validated as competitive in this task using intermediate quality spectra \citep[e.g.][]{Ghezzi(c)}.
     
     \subsection{Calibration stars} 
     \label{calibr}
     We observed a sample of 69 solar-type stars 
     for calibrating spectral indices.
     They are called hereafter ``calibration stars'' and are
     listed in Table~\ref{tab:cab_stars}.
     Their atmospheric parameters were extracted from the literature
     and the sources are provided in the table.
     Most of the sample (39 stars) is found in
     \citet{Ghezzi(a),Ghezzi(b)}, where \teff~determinations are  
     based on excitation \& ionization equilibrium of Fe lines.
     The rest of the stars belong to catalogs where \teff~was also derived by the same technique, 
     excepting 16 stars for which parameters were extracted from \citet{porto2014},
     where \teff~is the average of 
     photometric calibrations and H$\alpha$ line-profile fitting.
     The mean quoted precision of this sample is $\sim 40$~K, and 0.02~dex, 0.10~dex in \teff, [Fe/H], and \logg.
     10 stars were observed twice with the purpose of estimating uncertainties of the indices measurements: 
     HD 146233, HD 150248, HD 112164, HD 131117, HD 34721, HD 20029, HD 206395, HD 212330, 
     HD205420 and HD 215648. 
     
\begin{figure} 
     \centering
     \includegraphics[width=8.5cm]{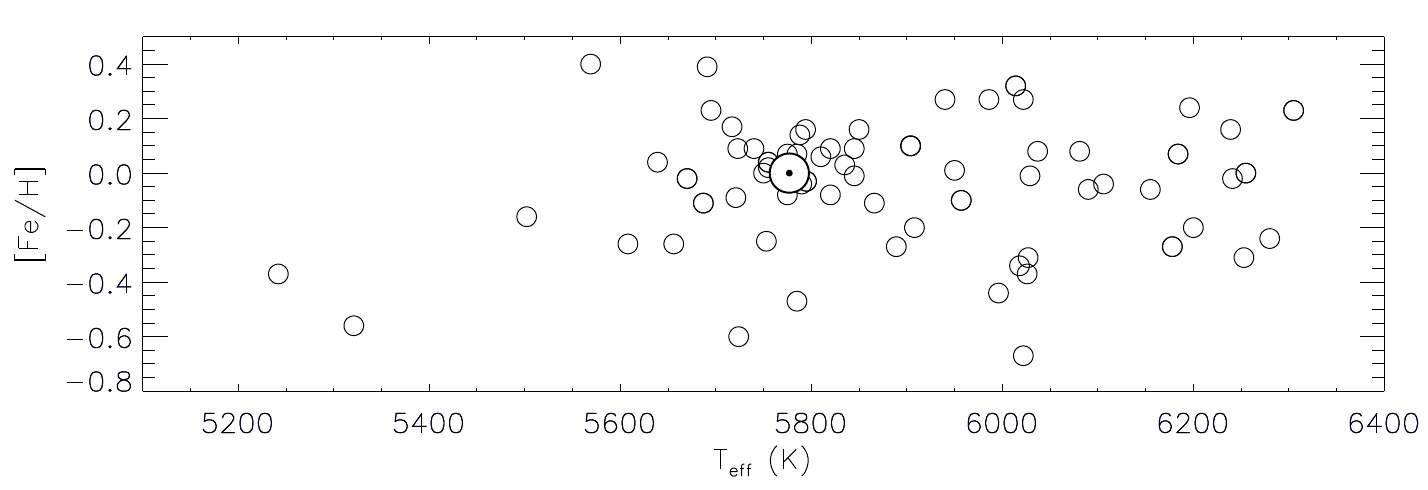}
     \includegraphics[width=8.5cm]{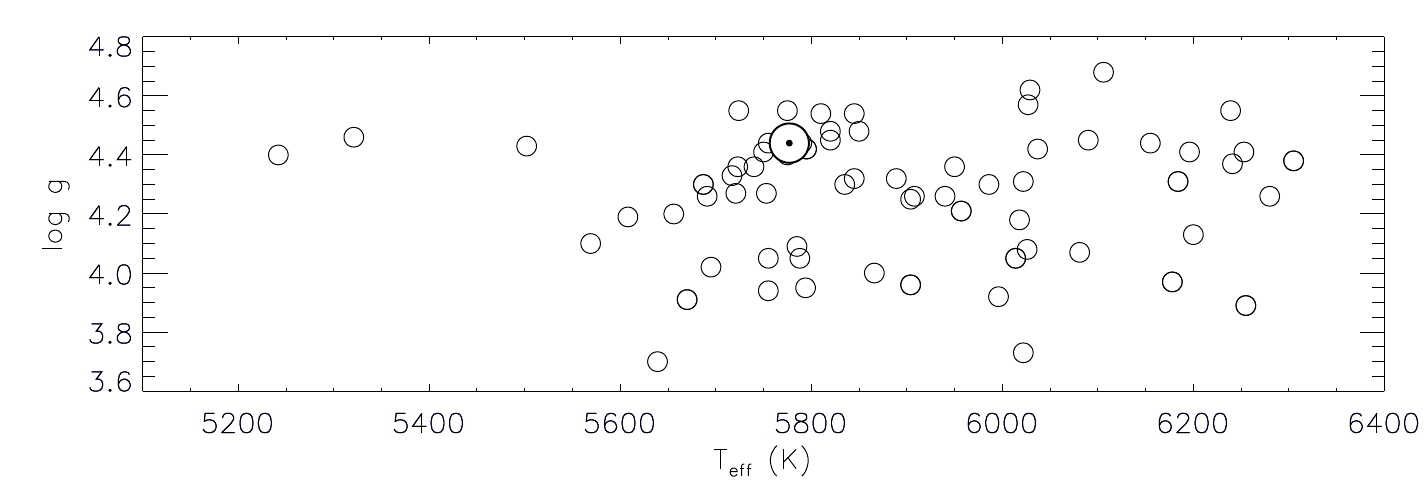}
     \includegraphics[width=8.5cm]{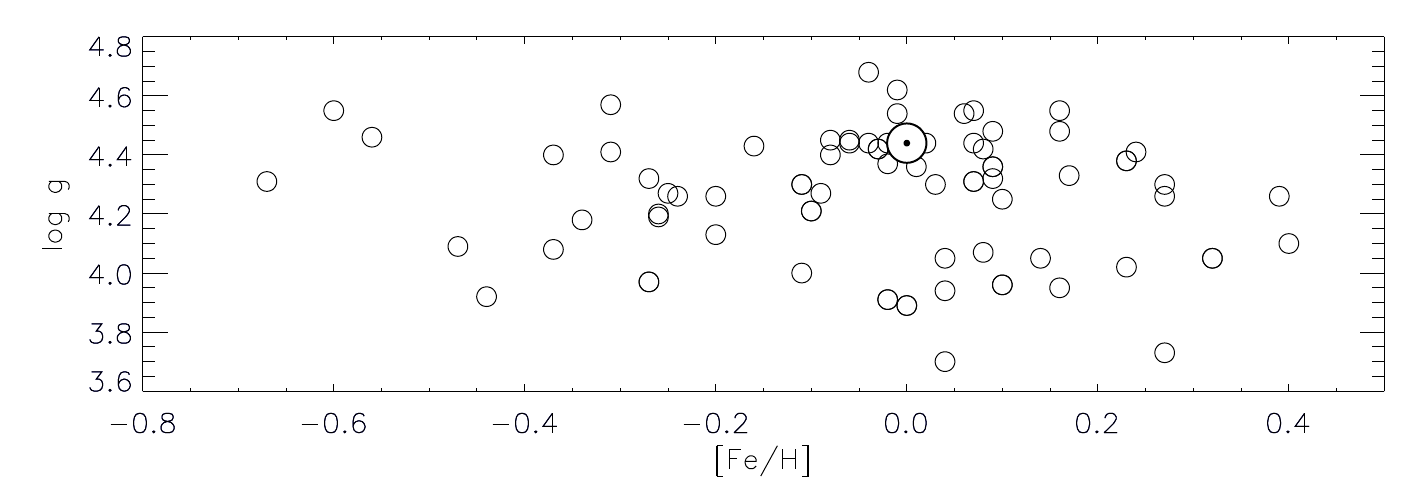}
     \caption{\label{Fig1} Distribution of the atmospheric parameters of the calibration stars around the solar values. The solar parameters are represented by the symbol $\odot$.}
     \end{figure}

     The distribution of the calibration stars in the parameter space is shown in Fig.~\ref{Fig1}.  
     They are more densely packed around the solar parameters
     \teff~= 5772~K \citep{Prsa2016,Heiter}, 
     [Fe/H] = 0~dex, 
     and $ \log{g} = 4.44$~dex 
     to calibrate as best as possible the solar analogs area.
     Notice that the area for \mbox{\teff~< 5600 K} is practically empty.
     This feature highlights the applicability limitation of our method, especially towards cooler and metal-poor stars.
     Therefore we choose to adopt the applicability range of our calibrations as follows: $5600\leq$ \teff~$\leq$ \mbox{6300 K}, 
     $-0.3 \leq$ [Fe/H] $\leq 0.4$~dex, \logg~$\geq4.1$~dex.
     
    \subsection{Identification of indices}
    Following \citep{Ghezzi(c)} we only selected indices dominated by iron peak elements, from both neutral and ionized species -- with a contribution of more than 90\%-- (Fe I, Fe II, Ti II, V I, Cr I, Cr II, Mn I, Co I, Ni I). These indices are shown to best correlate with atmospheric parameters. The inspection was carried out along the available spectral range avoiding the H$\alpha$ profile.
    
\begin{figure} 
     \centering
     \includegraphics[width=8.5cm]{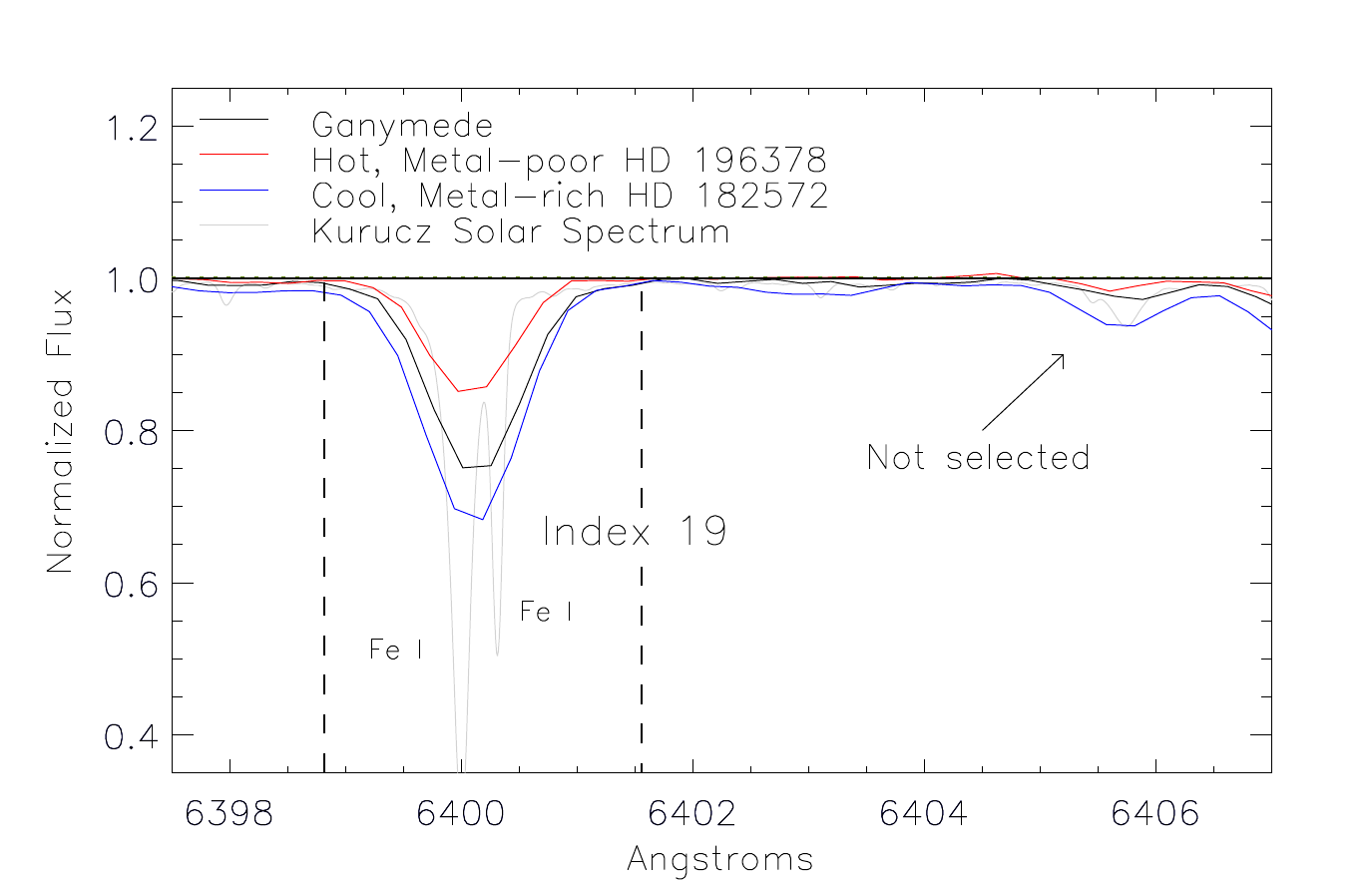}
     \caption{Definition of one spectral index. The KPNO atlas with resolution $R = 500~000$ is shown in gray. The three other spectra with resolution $R = 8000$ are from the Sun (black), HD~196378 (red), and HD~182572 (blue). The dashed lines mark the boundaries of index 19, formed by two Fe I lines.}
     \label{fig:selection} 
     \end{figure}

    Line identification was performed by comparing simultaneously the Kitt Peak National Observatory solar atlas \citep[KPNO,][]{kurucz2005}\footnote{\url{http://kurucz.harbard.edu/sun.html}} with spectra of the Sun (reflected off Ganymede), HD 19637, and HD 182572; as shown in Fig.~\ref{fig:selection}.
    The comparison between KPNO and Ganymede helps to visually identify metallic lines into the indices, whose contributions were estimated by their EWs as listed in the catalog of \citet{moore}.
    The element species were also checked using the VALD3 database \citep{vald3}.
    The spectrum of HD~19637 (hot and metal-poor star) was used for dismissing the weakest indices, while the spectrum of HD~182572 (cool and metal-rich star) was used to better define the wavelength limits of the indices. We selected 42 well defined indices that were submitted to the sensitivity test described below.
    
    \subsection{Calibration by PCA}
    Correlations of the EWs of the indices with the atmospheric parameters \teff, [Fe/H], and \logg~are approximated by a Taylor polynomial expansion of second order, as given in Eq.~\ref{eq:taylorexpan}.
        \begin{equation} \label{eq:taylorexpan}
    \begin{split}
    EW~(\text{m\AA}) &= c_0 + c_1 \text{[Fe/H]} + c_2 T_{\text{eff}} + c_3 \log{g}~+ \\
    & + c_4 \text{[Fe/H]} T_{\text{eff}} + c_5 \text{[Fe/H]}  \log{g}  + c_6 T_{\text{eff}} \log{g}~+ \\
    & + c_7 (\text{[Fe/H]})^2 + c_8 (T_{\text{eff}})^2 + c_9 (\log{g})^2
    \end{split}
    \end{equation}
    \noindent
    Following the same procedures of \citet{Ghezzi(c)}, we select 24 indices with the best sensitivity to \teff~and [Fe/H] (class 1 and 2, according to their definition), to which we then applied the PCA regression.
    
    %We specifically applied the principal component regression, which is a method based on the PCA. 
    The PCA extracts important information of correlated data sets, in which the direction of the greater variability of the correlations is searched. 
    It finds a new basis in which the data sets exhibit their greatest variance, 
    providing groups of non-correlated orthogonal components (Principal Components, PC's) based on linear combinations of the original input variables (the spectral indices EWs in our case). 
    This approach enables the extraction of the most relevant combinations of the original input variables and, thus, 
    they can be used for efficient discrimination of objects of different nature that present similar observables \citep[e.g.][]{blanco-cuaresma2015,hunt2012}, 
    and can be also calibrated against physically motivated variables, 
    such as \teff, [Fe/H], and \logg, as done by \citet{munoz2013}, and as we do in the present work.
    
    The variables were standardized to take into account their different scales as follows:
\begin{equation}
    Variable = \frac{Variable - \langle Variable \rangle}{\sigma (Variable)},
\end{equation}
where $ \langle Variable \rangle $ and $ \sigma (Variable) $ are, respectively, its average and standard deviation. 
We explored the correlations between the PC's and the atmospheric parameters of our calibration sample finding that the first (PC1) and the second (PC2) principal components are better related to all three parameters, i.e. they correspond to 90\% of the total cumulative variance of the data. The other, higher order, principal components do not show significant correlation with the atmospheric parameters and thus were discarded. We used the best regressive model to build a calibration for each one of the atmospheric parameters. Eqs. \ref{eq:tefPCA}, \ref{eq:fehPCA}, and \ref{eq:loggPCA}, show the atmospheric parameters as functions of PC1 and PC2:

\begin{equation}\label{eq:tefPCA}
    \begin{split}
        T_{\text{eff}} &= 5913(\pm 12) + 18(\pm 3)PC1 - 124(\pm 9)PC2\\&
        -7(\pm 3)(PC1 \times PC2)
    \end{split}
\end{equation}
\begin{equation}\label{eq:fehPCA}
    \begin{split}
        [\text{Fe/H}] &= -0.01(\pm0.01) - 0.039(\pm 0.0)PC1 \\ &- 0.042(\pm 0.0)PC2
    \end{split}
\end{equation}
\begin{equation}\label{eq:loggPCA}
    \begin{split}
        \log{g}  &= 4.30(\pm 0.03) - 0.0(\pm0.01)PC1 + 0.06(\pm 0.02)PC2 \\&-0.01(\pm 0.01) (PC1 \times PC2)
    \end{split}
\end{equation}
\noindent
The internal uncertainties of these analyses are 93~K, 0.06~dex, and~0.16 dex, for each atmospheric parameter, \teff, [Fe/H] and \logg, respectively.

%__________________________________________________________________
% ATMOSPHERIC PARAMETERS
%------------------------------------------------------------------
\section{Fundamental parameters of the candidates}
\label{sec:parameters}
\subsection{Spectroscopic effective temperature and metallicity}
We call hereafter spectroscopic parameters those derived from the PCA calibration of spectral indices, and we simbolize them hereafter by \teffPCA, [Fe/H]$^{PCA}$, and \logg$^{PCA}$.
The adopted values and uncertainties of stellar atmospheric parameters were estimated from $10^5$ Monte Carlo (MC) simulations, assuming that the EW's errors follow Gaussian distributions. 
The fractional EW errors estimated from the subsample of stars with two observations are found to be $\sim4$\% (the stars are indicated in Table~\ref{tab:cab_stars}). 
The outcome of MC simulations are EW distributions that were propagated by Eq.~\ref{eq:tefPCA}, \ref{eq:fehPCA}, and \ref{eq:loggPCA} to finally obtain a distribution of atmospheric parameters from which the most probable values and their errors were associated to the medians and standard deviations.

\begin{figure}
\includegraphics[width=8.0cm]{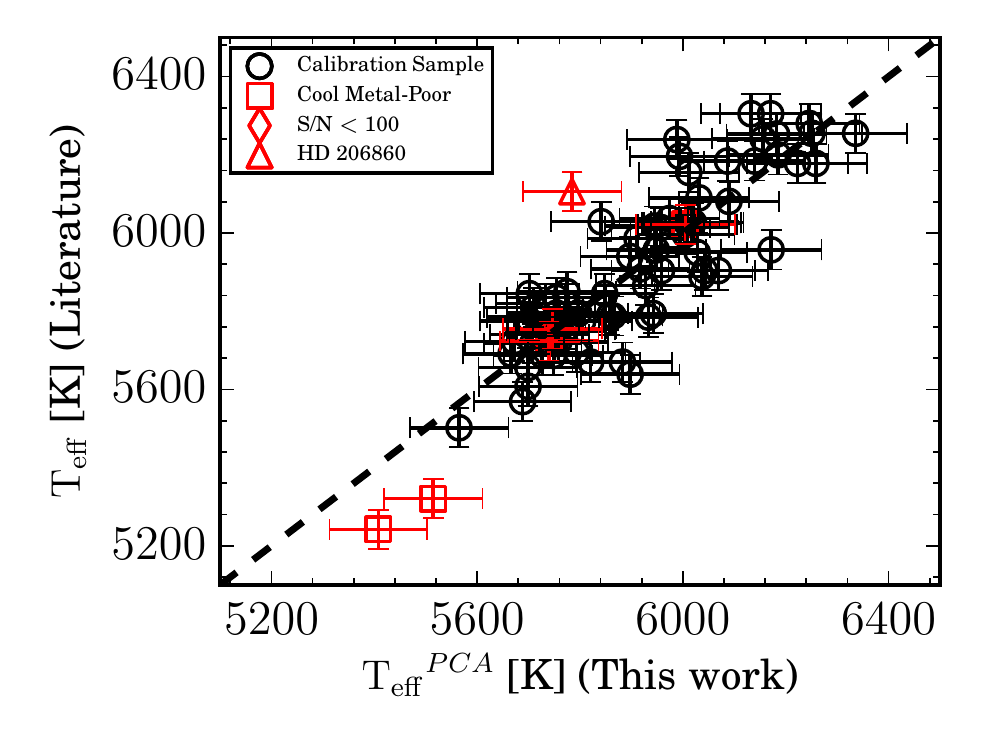}
\includegraphics[width=8.0cm]{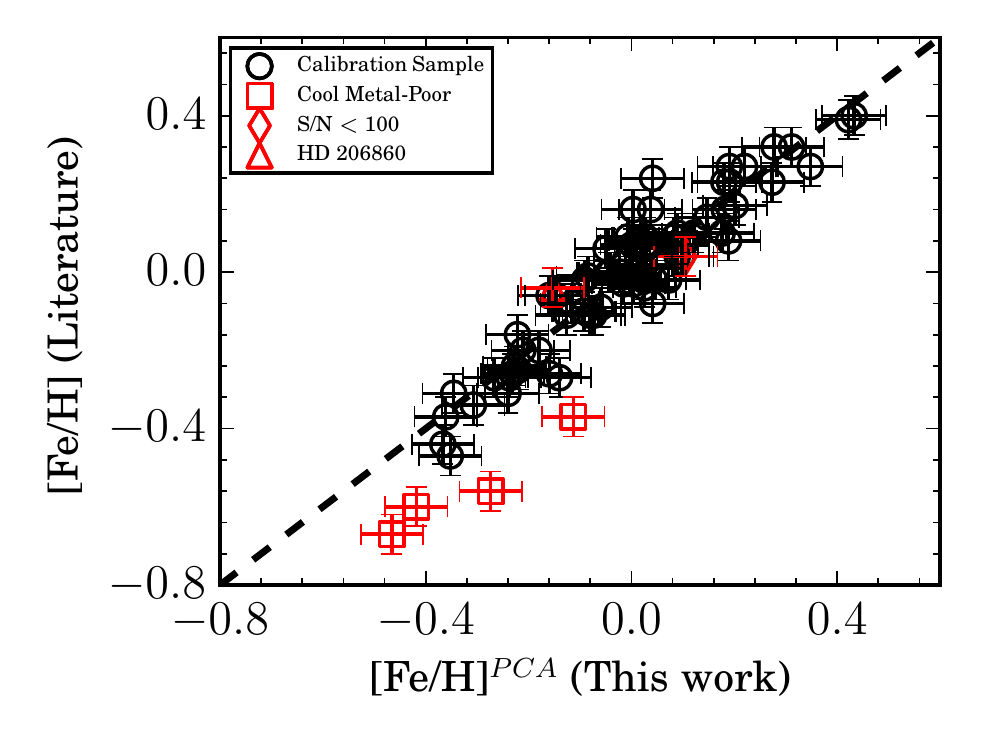}
\includegraphics[width=8.0cm]{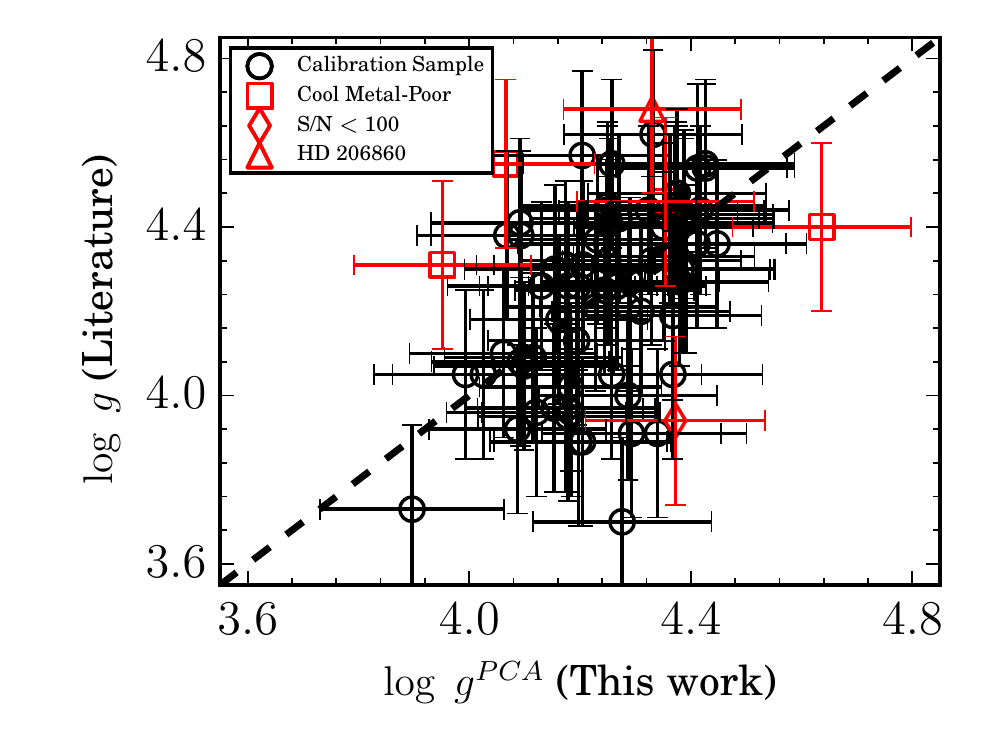}
\caption{Comparison between the atmospheric parameters from the literature and those from our $PCA$ calibration.}
\label{fig:paramPCA}
\end{figure}

We applied this procedure to the calibration sample in order to check the consistency between the PCA-based parameters and those of the literature, the results are shown in Fig.~\ref{fig:paramPCA}.
The agreement is satisfactory only for \teff~and [Fe/H].  
Accordingly, \logg~values derived by spectral indices are dismissed, and we determine them by evolutionary tracks in Sect.~\ref{subsec:logg}.
The plots confirm that the stars with parameters out of the applicability range are biased (red squares) to hotter and more metal-rich diagnostics, in general. 
The plots also show that the spectroscopic PCA parameters of the only calibration star with a spectrum of $S/N < 100$ (a value representative of the candidate star sample) agree with the literature values. 
Literature \teff's of the outlier HD~206860 (red triangle) were reviewed; the initially adopted \teff~was found to be too hot, being actually the hottest one in the published range.

The atmospheric parameters of the candidates derived by the procedures described above are presented in Table~\ref{tab:canddtstars}.
We keep henceforth the notation \teffPCA~for temperatures derived by spectral indices, and the values presented in the table were corrected by the equation given in Sect.~\ref{consistency}. Only parameters within the range of applicability pointed out in Sect.~\ref{calibr} are provided.

%TEMPERATURA PCA AMOSTRA CALIBRADORA
%__________________________________________________________________
%SUBSEC: PHOTOMETRIC EFFECTIVE TEMPERATURES
%------------------------------------------------------------------
\subsection{IRFM-photometric effective temperature} 
\label{phtemp}
We derived another set of temperatures using the metallicity-dependent color calibrations of \citet{Casagrande2010} based on the InfraRed Flux Method  \citep[IRFM][]{Blackwell1977,Blackwell1979,Blackwell1980}, we symbolize it henceforth as \teffP. 
Casagrande et al. corrected the systematics of previous IRFM implementations, 
their temperature scale was found to be in precise agreement with \teff~derived from interferometric measurements for the metallicity range in this work \citep{casagrande2014,giribaldi2018}.
Thus, we consider the \teffP~as the standard scale.

We derived \teffP~by computing the weighted mean 
of the temperatures obtained with the 
\mbox{$ (B-V) $}, \mbox{$ (B-V)^{Ty} $}, \mbox{$ (V-J) $}, \mbox{$ (V-H) $} and \mbox{$ (V-K_s) $} colors,
and [Fe/H]. 
The total uncertainty $\sigma$\teffP~was computed expanding the errors of colors, [Fe/H], and the internal uncertainty of the color calibration given by the authors.

\subsection{Consistency between spectroscopic and IRFM-photometric effective temperatures}
\label{consistency}
\begin{figure*}
\centering
\includegraphics[width=7cm]{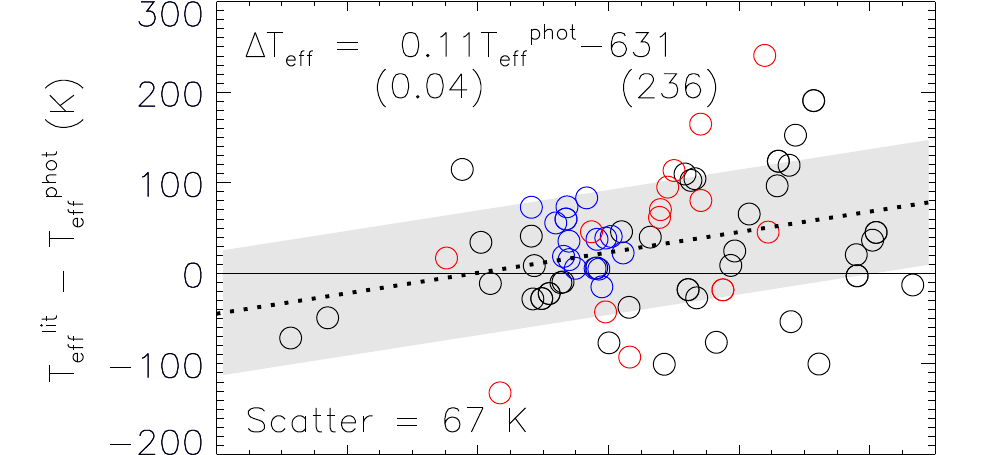}
\includegraphics[width=7cm]{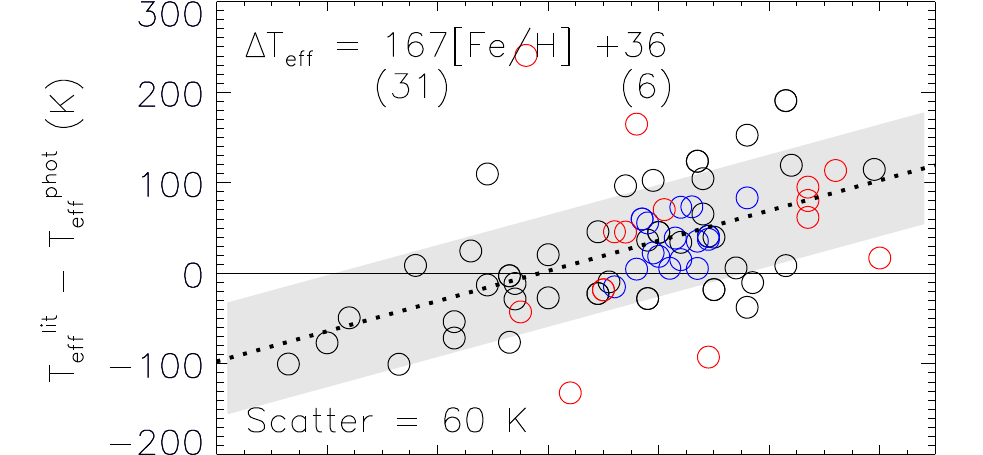}
\includegraphics[width=7cm]{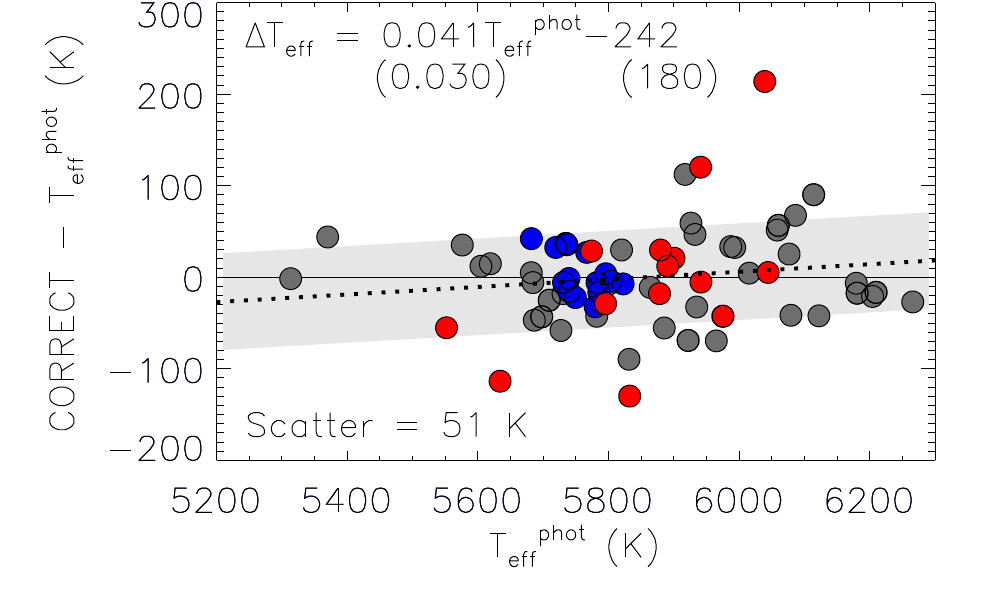}
\includegraphics[width=7cm]{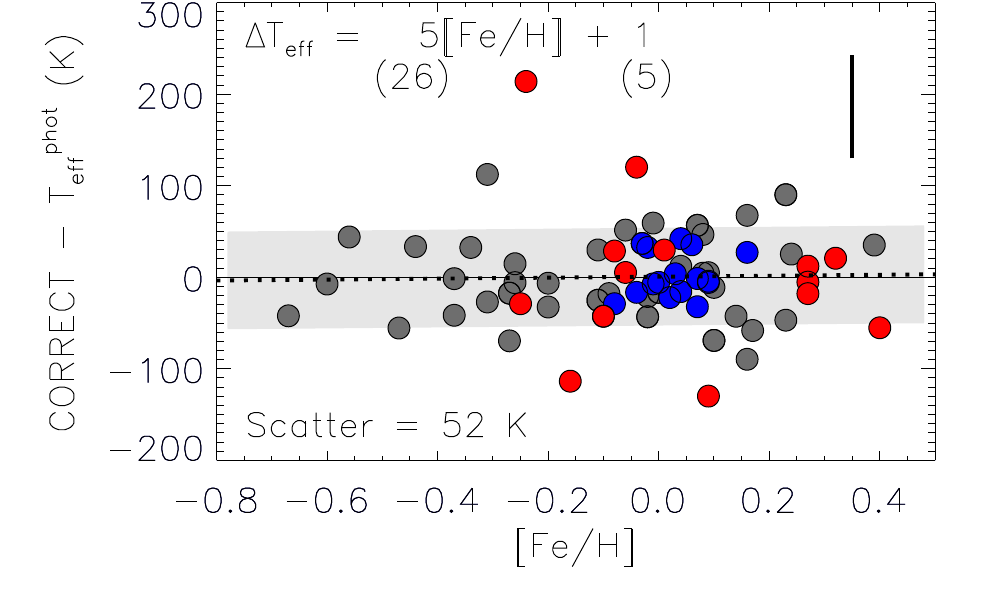}
\caption{\textit{Top panels: } Comparison between effective temperatures of the literature \tefflit~and \teffP of the calibration stars. Dark circles are the stars from \citet{Ghezzi(a),Ghezzi(b)}, blue circles are the stars from \citet{porto2014}, and red circles are the stars from all other sources listed in Table~\ref{tab:cab_stars}. 
The dotted lines and the shades are the trends and the 1$\sigma$ dispersion around them, respectively, whose equations are shown along with the error of their coefficients in brackets. \textit{Bottom panels: }Same as in the top panels but for \tefflit~corrected by \teffP~= (\tefflit~$+~410$)$/1.08 -153$[Fe/H] + 22. The vertical bar represents the mean uncertainties of the spectroscopic and photometric temperatures added: 40 + 73 K.}
\label{fig:agree}
\end{figure*}

\begin{figure*}
    \centering
    \includegraphics[width = 8.5cm]{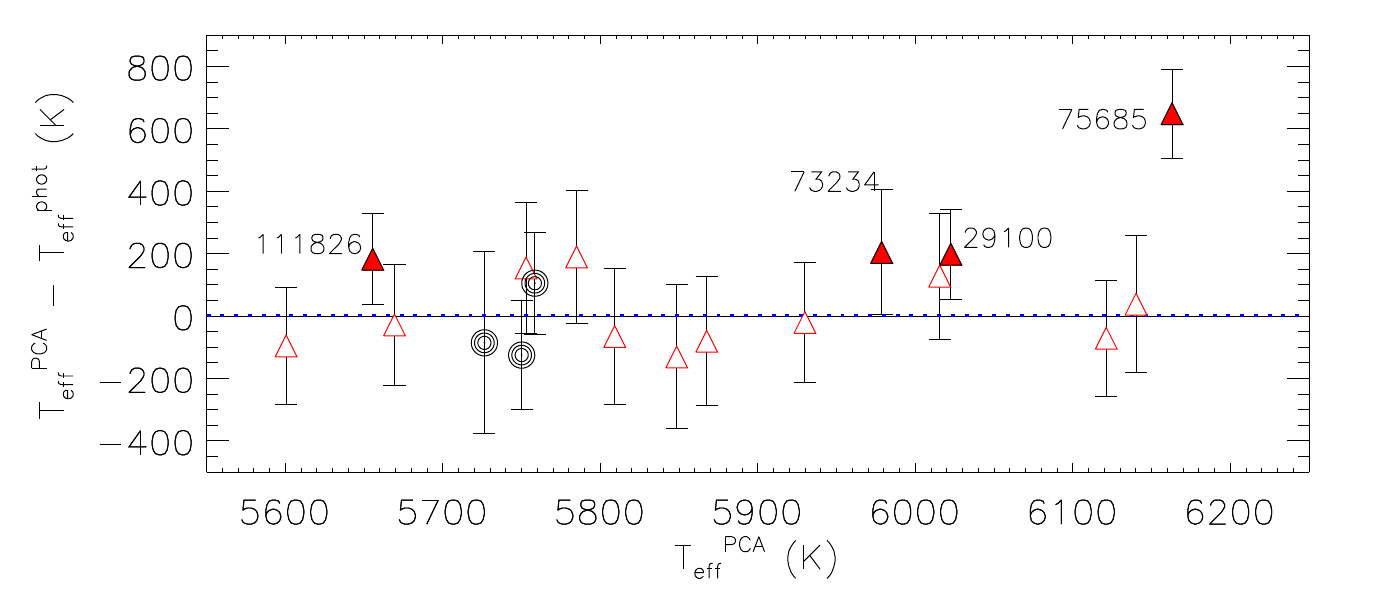}
    \includegraphics[width = 8.5cm]{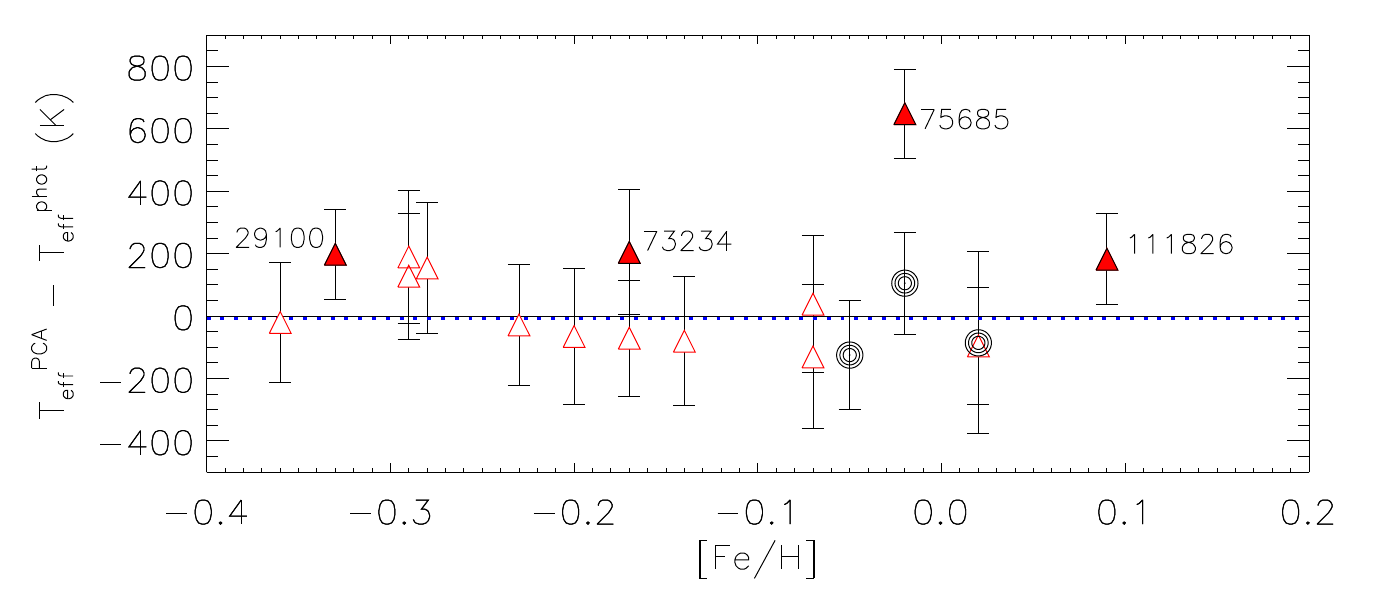}
    \caption{As in the bottom panels in Fig.~\ref{fig:agree} for the candidates, 
    but with \teffPCA~in the abscissa. Circles represent the stars in Table~\ref{tab:results} with \teff, 
    [Fe/H], and \logg~close to solar within 1$\sigma$ errors, while triangles represent all other stars. Filled symbols represent the stars with significant temperature differences, they are also labeled by their HIP number. The dotted blue lines at $-6$ K represent the average \teff~difference, computed for the unlabelled stars only.}
    \label{scales}
\end{figure*}
The accuracy of effective temperature measurements and the consistency between temperature scales is a recurrent topic in stellar astrophysics, and its importance increased with the discovery of exoplanets and the arrival of precise data from large surveys.
Spectroscopic and photometric scales show discrepancies for parameters far from solar, see for example comparisons in \citet{Casagrande2010}, \citet{Heiter}, and references in Table~\ref{tab:cab_stars}. 

Precise radius measurements from interferometry allow to derive \teff~semidirectly for nearby stars,
thus they can be used to test the accuracy of  model-dependent techniques, for example by using the
\textit{Gaia Benchmark Stars} \citep{Heiter}. 
This task was performed by \citet{giribaldi2018} for a parameter space similar to that analyzed in this work. 
They found that the IRFM \teff~scale implemented by \citet{Casagrande2010} agree with the interferometric one, as already reported by \citet{casagrande2014}. 
On the other hand, they found that spectroscopic \teff~scales based in LTE + 1D model atmospheres present a bias as a function of [Fe/H], 
producing \teff~underestimations/overestimations for metal-poor/metal-rich stars,
regardless of line lists and particular implementations of the technique.

Giribaldi et al. showed a trend of the spectroscopic \teff~scale of \citet{Ghezzi(a), Ghezzi(b)} (our calibrations are based mainly on it) with respect to \teff~based on interferometry as a function of [Fe/H],
and provided corrections for it, that is, to empyrically convert this scale to the interferometric one (or to the IRFM one, which is equivalent).
In Fig.~\ref{fig:agree}, we show the comparison between \teffP~and the temperatures from the literature (\tefflit), which are essentially spectroscopic,
for the calibration stars. 
We observe a similar trend to that shown by Giribaldi et al. (Fig.~10 in the paper), and find that the equations given by the authors 
above\footnote{Reduced into one equation here: \teffP~= (\tefflit~$+~410$)$/1.08 -153$[Fe/H] + 22; where \tefflit~represents the temperatures from the literature listed in Table~\ref{tab:cab_stars}.} subtract the trend.
This equation is applied to \teffPCA~of the candidates, so the values listed in Table~\ref{tab:results} are corrected values. These empirical corrections may be not elegant, but they are useful to assert accurate \teff~for non-solar [Fe/H].
They removed, for example, $\sim50$~K excess
for the hot more metal-rich candidates HIP~10663 and HIP~75685, but
they do not affect \teff~of solar analogs. 
\teffP~and \teffPCA~(corrected by the equation given above) of the candidates are compared in Fig.~\ref{scales}, 
where \teffPCA~is used as the absolute scale in the abscissa due to its insensitivity to reddening.
No trends are observed in the comparisons against \teff~and [Fe/H], and the offset between both scales is practically null (we considered only the stars without any evidence of reddening to compute this difference, see Sec.~\ref{sect:redd} for details). This asserts the consistency of the corrected spectroscopic and photometric scales.  
Stars with significant temperature differences are highlighted by filled symbols in Fig.~\ref{scales}. Their associated reddening values are discussed in Sect.~\ref{sect:redd}.

\subsection{Effective temperature from H$\alpha$ profiles}
Once the consistency between spectroscopic and photometric scales is realized by applying corrections to \teffPCA,  
significantly cooler \teffP~suggest the presence of reddening. 
Here we verify by means of H$\alpha$ profiles whether 
significant temperature differences are indeed due to reddening effects on \teffP.
Although the limited resolution of our spectra 
%suppress the windows free from metal-line contamination required to fit observed with synthetic profiles
does not allow a precise application of the H$\alpha$ profile fitting
and prevents the precise determination of \teff, temperature differences higher than $\sim200$ K are discernible. 

\teff~from H$\alpha$ is not affected by reddening and its determination practically does not depend on other parameters at solar metallicity \citep[e.g.][]{Fu1993, BPO2002}. Its main source of error is the normalization, which is a complex task in high resolution spectra, due to the short wavelength ranges that the profile leaves available into a spectral order for interpolating a polynomial that can reliably approximate the spectrograph response. However, our moderate resolution spectra are more than 3 times wider than the profile region, hence our normalization recovers the profile shape reasonably well.
In Fig.~\ref{fig:Halpha} we compare the observed profiles of HIP 67692 and HIP 75685 with synthetic profiles, from the grid of \citet{BPO2002}, corresponding to temperatures similar to their \teffPCA~and \teffP. This grid is found to be $-28$~K accurate for the metallicty of these stars \citep{giribaldi2018}. We also plot observed profiles of other candidates with \teffPCA values very similar to these two stars. The top plot in the figure shows that the profile of HIP~67692 is more compatible with its \teffP, while the bottom plot favors \teffPCA~for HIP~75685, whose profile is slightly deeper than that of HIP~10663 with \teffPCA~$\sim$ 6150~K.
Accordingly, \teffPCA $\sim$ 5400~K for HIP~67692 is not listed in Table~\ref{tab:results}, since this value lies out of the valid range of our indices calibration.

%__________________________________________________________________
%SUBSEC: SUPERFICIAL GRAVITIES, MASSES AND AGES
%-----------------------------------------------------------------
\subsection{Surface gravity, mass, and age}
\label{subsec:logg}
From the Gaia parallaxes \citep{Gai(c)}, plus the best \teff, and [Fe/H] values shown in Tables~\ref{tab:canddtstars} and~\ref{tab:results}, we calculated stellar luminosities using bolometric corrections from \citet{andrae2018} and extinction values from our reddening estimates in Table~\ref{tab:redval}. Surface gravity, mass, and age were obtained from theoretical evolutionary tracks of \citet{kim2002} and \citet{yi2003} following the procedure described in \citet{grieves2018}.

\begin{table*}
\centering
\normalsize
\caption{Fundamental parameters of the observed faint solar analog candidates. The first column lists the Hipparcos number. The three faint solar analogs with atmospheric parameters compatible with those of the Sun within 1$\sigma$ errors are highlighted in large bold numbers, and the other three within 2$\sigma$ errors in small bold numbers. 
The brackets indicate stars with suspected significant reddening values: the latter are listed in Table~\ref{tab:redval}.
Column 2 gives the best \teff~obtained by the weighted average 
of the values in columns 3 and 4.
Column 3 displays \teffPCA~corrected by the equation given in Sect.~\ref{consistency}. Column 4 displays \teffP~from the photometric calibrations of \protect{\citet{Casagrande2010}} and the colors in Table~\ref{tab:canddtstars} Columns 5 to the last are self-explanatory. Three candidates listed in table 1 but for which we could not derive atmospheric parameters from PCA because their values lie outside the valid range of our calibrations are not listed here: they are HIP~13964, HIP~24742 and HIP~69232.}
\label{tab:results}
\begin{tabular}{l c c c c c c c c}
\hline\hline 
HIP &  best \teff & \teffPCA~$\pm~97$ &  \teffP  &  [Fe/H]~$\pm~0.06$  & \logg & Mass & Radius & Age \\
 &  (K)  &  (K) &  (K)    & (dex) & (dex) & (M$_\sun$) & (R$_\sun$) & (Gyr)\\
\hline
\large{\textbf{991}} & $ 5829 \pm 85 $ & 5750 & $ 5875 \pm 74 $ & $ -0.05 $ & $ 4.38 \pm 0.06 $ & $0.99 \pm 0.05$ & $ 1.07 \pm 0.05 $ & $6.2 \pm 2.7$ \\
\large{\textbf{5811}} & $ 5653 \pm 67 $ & 5600 & $ 5696 \pm 88 $ & $ +0.02 $ & $ 4.39 \pm 0.05 $ & $0.96 \pm 0.05$ & $ 1.04 \pm 0.04 $ & $ 7.9 \pm 2.6 $ \\
6089 & $ 5684 \pm 20 $ & 5669 & $ 5698 \pm 95 $ & $ -0.23 $ & $ 4.46 \pm 0.04 $ & $0.90 \pm 0.04$ & $ 0.93 \pm 0.03 $ & $ 6.9 \pm 1.5 $ \\
8853 & $ 6160 \pm 50 $ & 6121 & $ 6192 \pm 87 $ & $ -0.17 $ & $ 4.32 \pm 0.04 $ & $1.09 \pm 0.05$ & $ 1.20 \pm 0.04 $ & $ 4.10 \pm 1.0 $ \\
10663 & $ 6125 \pm 26 $ & 6140 & $ 6102 \pm 120 $ & $ -0.07 $ & $ 4.05\pm 0.03 $ & $1.24 \pm 0.05$ & $ 1.74 \pm 0.04 $ & $ 4.4 \pm 0.6 $ \\
18941 & $ 5955 \pm 90 $ & 6015 & $ 5887 \pm 103 $ & $ -0.29 $ & $ 4.39 \pm 0.07 $ & $0.96 \pm 0.06$ & $ 1.04 \pm 0.04 $ & $ 6.4 \pm 2.9 $ \\
$[$29100$]$ & \teffPCA & 6022 & $ 5824 \pm 46 $ & $ -0.33 $ & $ 4.49 \pm 0.05 $ & $0.95 \pm 0.05$ & $ 0.92 \pm 0.04 $ & $ 3.3 \pm 2.3 $ \\
31845 & $ 5705 \pm 132 $ & 5785 & $ 5596 \pm 113 $ & $ -0.29 $ & $ 4.47 \pm 0.06 $ & $0.89 \pm 0.05$ & $ 0.91 \pm 0.04 $ & $ 7.2 \pm 4.2 $ \\
48272 & $ 5941 \pm 14 $ & 5930 & $ 5950 \pm 92 $ & $ -0.36 $ & $ 4.40 \pm 0.04 $ & $0.93 \pm 0.05$ & $ 1.01 \pm 0.04 $ & $ 6.8 \pm 0.9 $ \\
\textbf{55619} & $ 5686 \pm 69 $ & 5758 & $ 5653 \pm 65 $ & $ -0.02 $ & $ 4.37 \pm 0.05 $ & $0.96 \pm 0.05$ & $ 1.06 \pm 0.04 $ & $ 8.0 \pm 2.6 $ \\
56870 & $ 5687 \pm 108 $ & 5753 & $ 5599 \pm 112 $ & $ -0.28 $ & $ 4.49 \pm 0.05 $ & $0.89 \pm 0.05$ & $ 0.89 \pm 0.04 $ & $ 6.4 \pm 4.0 $ \\
\textbf{61835} & $ 5895 \pm 88 $ & 5848 & $ 5979 \pm 132 $ & $ -0.07 $ & $ 4.30 \pm 0.05 $ & $1.02 \pm 0.04$ & $ 1.19 \pm 0.05 $ & $ 7.1 \pm 2.4 $ \\
67692* & --- & --- & $ 5427 \pm 37 $ & $ -0.04 $ & $ 3.83 \pm 0.04 $ & $1.23 \pm 0.04$ & $ 2.25 \pm 0.06 $ & $ 5.2 \pm 0.4 $ \\
\large{\textbf{69477}} & $ 5744 \pm 49 $ & 5726 & $ 5812 \pm 193 $ & $ +0.02 $ & $ 4.46 \pm 0.05 $ & $0.99 \pm 0.05$ & $ 0.98 \pm 0.04 $ & $ 4.1 \pm 2.7 $ \\
$[$73234$]$ & \teffPCA & 5979 & $ 5775 \pm 101 $ & $ -0.17 $ & $ 4.28 \pm 0.13 $ & $1.04 \pm 0.06$ & $ 1.22 \pm 0.16 $ & $ 5.8 \pm 2.0 $ \\
$[$75685$]$ & \teffPCA & 6163 & $ 5515 \pm 44 $ & $ -0.02 $ & $ 4.40 \pm 0.04 $ & $1.10 \pm 0.04$ & $ 1.10 \pm 0.04 $ & $ 2.1 \pm 1.5 $ \\
107605 & $ 5835 \pm 45 $ & 5809 & $ 5874 \pm 120 $ & $ -0.20 $ & $ 4.30 \pm 0.03 $ & $0.95 \pm 0.04$ & $ 1.14 \pm 0.04 $ & $ 0.95 \pm 1.8 $ \\
\textbf{$[$111826$]$} & \teffPCA & 5655 & $ 5474 \pm 47 $ & $ +0.09 $ & $ 4.45 \pm 0.06 $ & $0.98 \pm 0.05$ & $ 0.98 \pm 0.04 $ & $ 5.0 \pm 3.1 $ \\
\hline
\multicolumn{9}{l}{\small*\teffP~and \teff~from H$\alpha$ agree for this candidate. These values are out of the PCA applicability range, and \teffPCA~ was found to be significantly} \\
\multicolumn{9}{l}{\small hotter than \teffP and \teff~from H$\alpha$.
No reddening was estimated for this candidate and we consider its atmospheric parameters as unreliable.}
\end{tabular}
\end{table*}

\section{Reddening} 
\label{sect:redd}
The hunt for solar analogs begins, necessarily, by selecting candidates with solar photometric colors, as they are the most direct observational parameters able to quantify similarities between stars. 
Once photometric mimics to solar produced by the degeneracy of the atmospheric parameters (principally \teff--[Fe/H]) are identified and discarded, solar-like colors should lead to stars with the same atmospheric parameters as the Sun. 
However, in the presence of interstellar extinction a star which presents observed reddened colors equal to solar will have different combinations of atmospheric parameters tending to be hotter than the Sun. 
For hunters of solar analogs and twins at large distances, this implies that reddening corrections must be considered. 
For users of solar proxies, it also means that regardless of whether the intrinsic atmospheric parameters of a star are solar, the observed colors will always be reddened, i.e. $B-V_\star$ > $B-V_\sun$. 
Therefore, a faint star with apparent solar colors will have a flux distribution different from solar, 
and when used to remove the solar spectral signature from the spectrum of the target, it will introduce systematic trends in its spectral albedo.
    
We estimate reddening values for the candidates for which \teffP~are smaller than \teffPCA. These are are shown as filled triangles in Fig.~\ref{scales}. The color excess $E(B-V) = (B-V)_{reddened} - (B-V)_{intrinsic}$ is then computed by the
difference between the color required to obtain the average \teffP~and that required to obtain \teffPCA~by using the calibrations of \citet{Casagrande2010}.
Since \teffP~was determined by the weighted average of several colors, $(B-V)_{reddened}$ are not exactly the same as those in Table~\ref{tab:canddtstars}.
Table~\ref{tab:redval} shows $E(B-V)$ estimated by this method: they can be considered as lower limit estimates of the actual reddening in $B-V$, since this actual reddening is somewhat diluted by the process of determining the average \teffP~also employing colors which are less affected by reddening than $B-V$.

\subsection{Extinction models}
We compare here our $E(B-V)$ estimates with those predicted by two extinction maps. Reddening estimations by other methods such as Ca II H \& K lines, Na I D lines \citep[e.g.][]{Brito, curtis2017}, and diffuse interstellar bands \citep{law2017} were not possible due to the limitations established by the resolution and wavelength coverage of our spectra.
The description of dust distribution in our Galaxy has progressed a lot over the last two decades for both 2D and 3D maps and models \citep{Robin2015, Sale2015}.
\citet{SFD} (hereafter SFD) published 2D maps based on the FIR emission detected by COBE/DIRBE satellite. This model was reviewed by \citet{Beers} in order to correct overestimations of the total reddening in internal regions of the Galaxy (hereafter SFD-B).

\citet[][hereafter A\&L]{AL2005} presented two models for interstellar extinction in the Galaxy that take into account the gas distribution for HI and HII. In the first model, the Galaxy is axisymmetric (ALA) and extinction increases linearly as function of distance. In the second model (ALS), the spiral structure is considered and the extinction increases by steps each time a spiral arm is crossed. They compared their models for a wide range of distances and directions by using some catalogues, such as \citet{Neckel1980}, \citet{Savage1985}, and \citet{guarinos1992}. The last catalogue has the majority of their stars located at distances up to 500 pc.

A\&L, \citet{Arce1999}, among other works find that \citet{SFD} overestimates extinction for $E(B-V) \sim 0.15$ mag. Some simplifications done in the map such as resolution and the unique value used for dust temperature are provided as explanations for it. The overestimations are expected to mainly affect the Galactic plane and towards molecular clouds, however it is not explored from which distance they start to be relevant. 

\begin{figure}[ht]
      \centering
      \includegraphics[width=8.5cm]{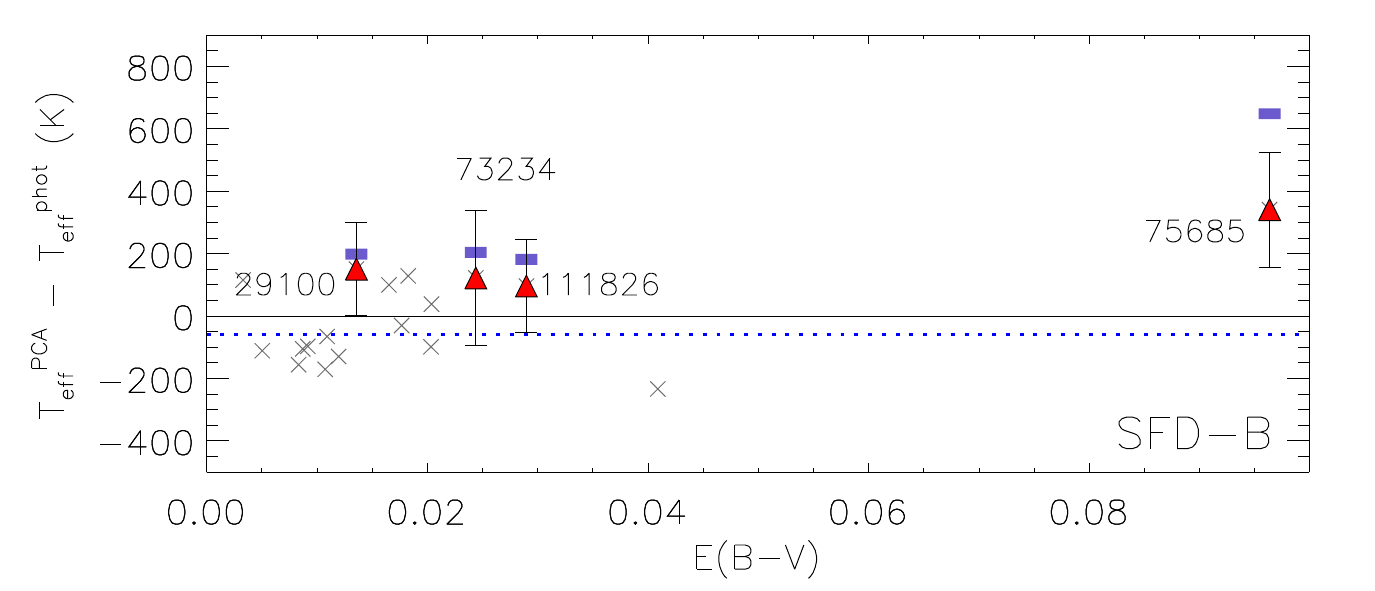}
      \includegraphics[width=8.5cm]{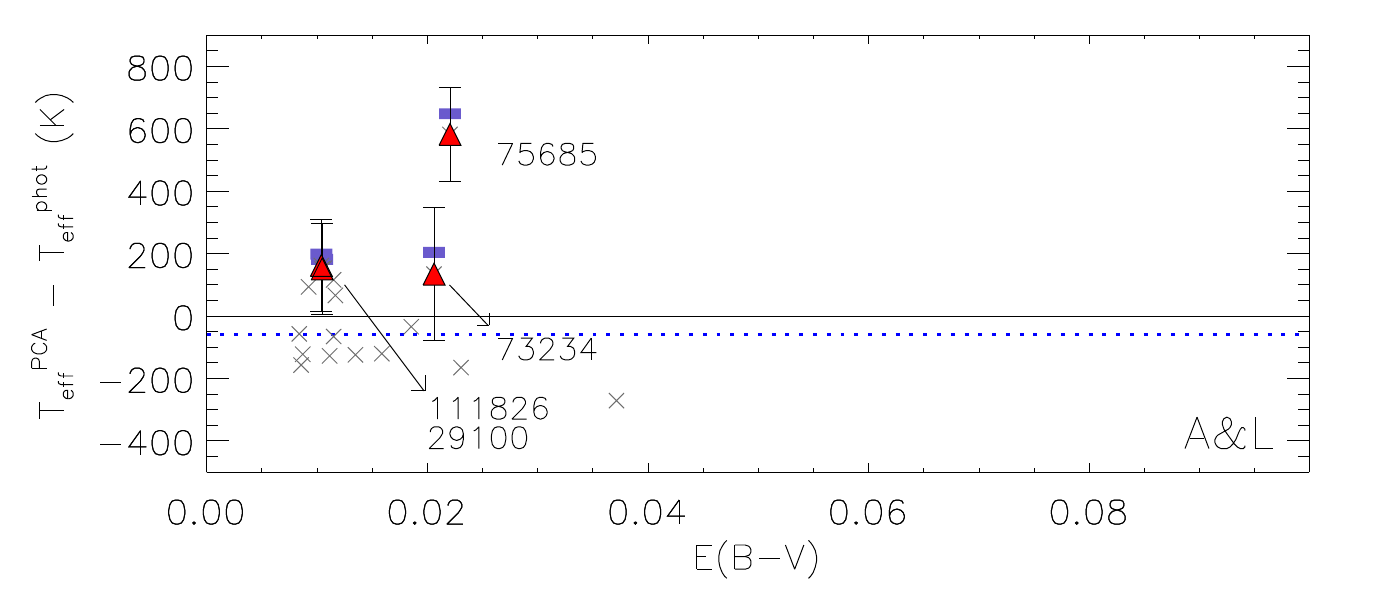}
      \caption{Difference between spectroscopic \teffPCA~and derredened \teffP~of candidates according to the extinction models by SFD-B (top panel) and A\&L (bottom panel).
      The stars with no significant differences are represented by crosses, and the others by the same symbols as in Fig.~\ref{scales}. As a reference, temperature differences from underredened colors are pointed by blue symbols, i.e. same values as in Fig.~\ref{scales}. In both plots, the dotted lines at $\sim -50$ K represent the average \teffPCA~$-$ \teffP~of the stars represented by the crosses.}
      \label{correlations}
      \end{figure}

Our choice was to use the ALA model of A\&L and the SFD-B model to test their consistency with our estimates from \teffPCA $-$ \teffP~at $\sim 170$ pc, their $E(B-V)$ are listed in Table~\ref{tab:redval}. 
Fig.~\ref{correlations} shows $E(B-V)$ from A\&L and SFD-B for all candidates plotted against \teffPCA $-$ \teffP~from derredened colors (red triangles), and also from non-derredened colors (blue bars) for the reader to check the corresponding temperature corrections.
$E(B-V)$, $E(B-V)^{Ty}$, $E(V-J)$, $E(V-H)$, and $E(V-K)$ were considered to obtain derredened \teffP; same values were used for Johnson and Thycho, while 2MASS reddenings were converted from Johnson by the relations given by \cite{zagury2012} for $R_V = 3.14$. The errors of derredened \teffP~were estimated expanding those of $E(B-V)$ given by the models, colors, parallax, [Fe/H], and photometric calibrations. These errors turned to be practically the same as those from non-derredened \teffP~because the error budget is dominated by parallax errors, which are negligible for the Gaia data in our distance range.  

Both models remove (or at least minimize) the differences of the labeled stars, except for HIP 75685. For this case, SFD-B predicts a substantially higher $E(B-V)$ than A\&L, but still lower than our estimate $E(B-V)$~$\sim$~0.20 mag. This value agrees with that of SFD, which is the total reddening predicted by the model for the line of sight, although it is in the range where the model predictions are known to present problems ($E(B-V) > 0.15$ mag) as pointed out above. Given the reasonable agreement between these independent estimates of reddening, we consider the case for these three objects as substantial, particularly for HIP~75685.

\begin{table}  
\caption{Color excess values $E(B-V)$ of candidates with significant \teffP~underestimations, compared with predictions from the SFD, SFD-B, and A\&L models plus our own estimate in column 4. The last column gives the \teffP~underestimation with respect to \teffPCA~implicated by the values in column 4.}
\small
\centering
    \begin{tabular}{c c c c c c}
    \hline \hline
         HIP & SFD & SFD-B & A\&L & this work & $\Delta$\teff~(K) \\
         \hline
         29100 & 0.0426 & 0.0136 & 0.0104 & $0.058 \pm 0.042$ & 198\\
         73234 & 0.0244 & 0.0244 & 0.0206 & $0.062 \pm 0.061$ & 204\\
         75685 & 0.1647 & 0.0963 & 0.0221 & $0.203 \pm 0.043$ & 648 \\
         111826 & 0.0290 & 0.0290 & 0.0104 & $0.063 \pm 0.051$ & 182\\
         \hline \hline
         \end{tabular}
    \label{tab:redval}
\end{table}

\section{Best faint solar analogs}
\label{best}
The results of the previous sections point towards the identification of a sample of faint solar analogs of $V \sim 10.5$ which reproduce well the atmospheric parameters of the Sun and should be good matches for its spectrophotometric flux distribution for a wide range of wavelengths. 
Three stars have atmospheric parameters agreeing with solar within 1$\sigma$ of their formal errors: HIP~991, HIP~5811 and HIP~69477. 
Other two candidates agree in the same sense but within 2$\sigma$ of their errors: HIP~55619 and HIP~61835. Their ages are found to be comparable or larger than the Sun's, and moreover their H$\alpha$ line cores do not show any discernible fill-in from a high level of chromospheric activity, which should be apparent even in moderately low resolution spectra \citep{lyraportodemello2005}. All evidence point to their being middle-aged, inactive solar analogs.
Their estimated masses and radii also closely match the solar ones within formal uncertainties, but HIP 61835 which has a slightly larger radius. 
They are reasonably well distributed across the sky but slightly biased towards southern declinations due to the reach of our observations.

Six additional objects have \teff~matching the solar one but appear as slightly metal$-$poor, in the $-$0.30 $<$ [Fe/H] $<$ $-$0.20 range. They are probably poorer solar matches for shorter wavelengths but should reproduce the Sun increasingly better towards redder spectral ranges and are probably very good in the infrared \citep{porto2014}. These are HIP~6089, HIP~18941, HIP~31845, HIP~48272, HIP~56870 and HIP~107605. Consistently with their more diverse atmospheric parameters, their masses and radii do not match the Sun's as closely as the best analogs, but all of them (excepting HIP~107605) appear to be old stars and thus free from a high degree of chromospheric activity, and are also reasonably well scattered in the sky. These latter stars may be considered by potential users to be reasonable matches to the Sun as a function of the desired precision and accuracy for the target observations.
All of the aforementioned eleven faint solar analogs are free from any evidence of reddening according to our analysis. 
We could not find any sign of binarity in the spectra of the candidates analysed, and better quality observations should be used to eliminate this possibility.
      
\section{Conclusions}
\label{sec:conclusions}

Motivated by the demand for faint spectrophotometric solar analogs, 
we implemented the methodologies to derive 
atmospheric parameters with optimized precision from moderately low resolution and S/N spectra.
We selected a sample of candidates with $V \sim 10.5$ in the Hipparcos catalog by matching the solar M$_V$ and $B-V$ values in the Johnson and Tycho systems, subsequently we submitted a subsample of them to spectroscopic analysis. 
The method for deriving atmospheric parameters consist on a system of 24 spectral indices, whose sensitivity to \teff~and [Fe/H] were mathematically modeled by the PCA regression.
The models were based on published spectroscopic \teff's (based on the excitation and ionization equilibrium of Fe lines in LTE + 1D model atmospheres), thus \teff~derived by the spectral indices may be also deemed spectroscopic. 
Considering the discrepancies between \teff~scales from different techniques at parameters far from solar, we assured the consistency of the spectroscopic \teff~with the photometric \teff~\citep{Casagrande2010} -- which is consistent with the interferometric \teff~of the \textit{Gaia Benchmark Stars} \citep[][]{Heiter} -- using the relations given by \citet{giribaldi2018}. The corrected spectroscopic \teff~are shown to match the photometric ones. 
Excepting for the stars showing evidence of reddening, finally adopted \teff~were derived by averaging the photometric and spectroscopic determinations. 
The finally derived spectroscopic \teff~and [Fe/H] have internal precision, respectively, of 97~K and 0.06~dex. 
The PCA index system is very successful in recovering atmospheric parameters with good precision, even for low S/N spectra, and may be used to study fainter stars in large databases; the accuracy of the such parameters entirely relies on the accuracy of the calibrating sample.

Surface gravities, masses, radii and ages were derived from the finally adopted atmospheric parameters and Gaia parallaxes by means of theoretical evolutionary tracks and isochrones \citep{kim2002, yi2003}. We identified 11 solar analogs to different degrees of resemblance to the Sun: their individual suitability as solar surrogates is judged in Sect.\ref{best}.
Fundamental parameters for them and other candidate stars that did not fully meet the requirements as solar analogs are displayed in Table~\ref{tab:results}; their photometric and astrometric parameters are listed in Table~\ref{tab:canddtstars}.

Initial candidates lie between 90 pc and 290 pc, and we estimated reddening for them independently from published extinction models, by comparing photometric \teff~with corrected spectroscopic \teff, since corrected spectroscopic \teff~are shown to be consistent with photometric ones. 
We find evidence of significant reddening for four candidates which present significant cooler photometric \teff. 
A common reddening value at these distances resulted in $E(B-V)$ $\sim0.06$ mag, which translates to a \mbox{$\sim200$ K} decrease in photometric \teff. 
Our estimates are validated by predictions from the SFD-B and A\&L extinction models, except for one star HD~75685, which appears to lie in a very dense region.

The identified analogs have no evidence of reddening, and may be used photometrically and spectroscopically for subtracting the solar signature with good precision from observations of Solar System bodies. 
In the visible and infrared regions they should present very good matching to the Sun, even in the UV up to 4000~\AA. 
Our reddening analysis shows that solar analog candidates will be progressively more affected by reddening. These stars will present spectra and colors that appear to belong to cooler stars as they become fainter (or more affected by reddening), as seems to be the case of HIP~75685. 
As future generations of larger telescopes increase the demand for faint stars matching the solar spectra, this will become a relevant issue to be addressed for the very faint solar analogs: photometrically selected solar analogs will not match the actual spectroscopic properties of the Sun.
      
\begin{acknowledgements}
    REG acknowledges scholarships from CAPES and ESO, GFPM acknowledges grant 474972/2009-7 from CNPq/Brazil, DLS acknowledges a scholarship from CAPES and FAPESP 2016/20667-8, and MLUM acknowledges scholarships from FAPERJ and CAPES. We thank the staff of the OPD/LNA for considerable support in the many observing runs carried out during this project. 
    We also thank the anonymous referee for helpful and constructive criticism to this work.
    Use was made of the Simbad database, operated at the CDS, Strasbourg, France, and of NASA’s Astrophysics Data System Bibliographic Services. 
    This publication makes use of data products from the Two Micron
    All Sky Survey, which is a joint project of the University of
    Massachusetts and the Infrared Processing and Analysis
    Center/California Institute of Technology, funded by the National Aeronautics and Space Administration and the National Science Foundation.
    This work presents results from the European Space Agency (ESA)
    space mission Gaia. Gaia data are being processed by the Gaia Data Processing and Analysis Consortium (DPAC). Funding for the DPAC is provided by national institutions, in particular the institutions participating in the Gaia MultiLateral Agreement (MLA). The Gaia mission website is \url{https://www.cosmos.esa.int/gaia}. The Gaia archive website is \url{https://archives.esac.esa.int/gaia}.
\end{acknowledgements}
    
\bibliographystyle{aa} % style aa.bst
\bibliography{riano}

\begin{appendix} %First online appendix
\newpage
\onecolumn
\section{ }
\begin{table}[b]
\caption{Atmospheric parameters of the calibration stars. The first column lists the HD number, except for the last star.
Columns 2, 3, and 4 are the atmospheric parameters, followed by column 5, with the literature source coded by number: 
1 \protect\cite{Ghezzi(a),Ghezzi(b)}, 2 \protect\cite{porto2014}, 
3 \protect\cite{daSilva2012}, 4 \protect\cite{daSilva2011}, 5 \protect\cite{daSilva(b)}, 
6 \protect\cite{Bensby}, 7 \protect\cite{Luck}, 8 \protect\cite{lyraportodemello2005}. The last column gives the $S/N$ --- both values are presented in case of two observations for one star.}
\label{tab:cab_stars}
\centering
\scriptsize 
\begin{tabular}{lcrccc}
\hline\hline
HD & $ T_{\mathrm{eff}} $ (K) & [Fe/H] & $ \log{g} $ & Author & $S/N$ \\
%	& (K)	&	 &	&	&    \\
\hline
1461 & 5717 & 0.17 & 4.33 & 1 & 155\\
1581 & 5908 & $-0.20$ & 4.26 & 1 & 166\\
2151 & 5866 & $-0.11$ & 4.00 & 1 & 324\\
4391 & 5829 & $-0.08$ & 4.45 & 8 & 201\\
7570 & 6196 & 0.24 & 4.41 & 1 & 297\\
8291 & 5835 & 0.03 & 4.30 & 2 & 141\\
9562 & 5794 & 0.16 & 3.95 & 1 & 217\\
9986 & 5820 & 0.09 & 4.48 & 2 & 297\\
10647 & 6155 & $-0.06$ & 4.44 & 1 & 223\\
10700 & 5321 & $-0.56$ & 4.46 & 1 & 471\\
12264 & 5810 & 0.06 & 4.54 & 2 & 194\\
16417 & 5788 & 0.14 & 4.05 & 1 & 272\\
17051 & 6239 & 0.16 & 4.55 & 1 & 269\\
19994 & 6081 & 0.08 & 4.07 & 1 & 192\\
20010 & 6280 & $-0.02$ & 4.26 & 7 & 368\\
20029 & 6184 & 0.07 & 4.31 & 1 & 224, 170\\
20630 & 5723 & 0.09 & 4.36 & 1 & 274\\
30495 & 5740 & 0.09 & 4.36 & 5 & 237\\
30562 & 5986 & 0.27 & 4.30 & 5 & 424\\
34721 & 5957 & $-0.10$ & 4.21 & 5 & 177, 252\\
36553 & 6022 & 0.27 & 3.73 & 5 & 498\\
39091 & 6037 & 0.08 & 4.42 & 1 & 207\\
39587 & 6029 & $-0.01$ & 4.62 & 1 & 426\\
43587 & 5950 & 0.01 & 4.36 & 5 & 109\\
43947 & 5889 & $-0.27$ & 4.32 & 1 & 117\\
52298 & 6253 & $-0.31$ & 4.41 & 1 & 204\\
65907 & 6027 & $-0.31$ & 4.57 & 1 & 320\\
98649 & 5775 & $-0.02$ & 4.44 & 2 & 151\\
105901 & 5845 & $-0.01$ & 4.54 & 2 & 117\\
112164 & 6014 & 0.32 & 4.05 & 3 & 131, 228\\
115382 & 5775 & $-0.08$ & 4.40 & 2 & 106\\
117939 & 5608 & $-0.26$ & 4.19 & 1 & 230\\
118598 & 5755 & 0.02 & 4.44 & 2 & 169\\
131117 & 5904 & 0.10 & 3.96 & 1 & 135, 184\\
134060 & 5904 & 0.10 & 4.25 & 1 & 149\\
138573 & 5750 & 0.00 & 4.41 & 2 & 294\\
146233 & 5795 & $-0.03$ & 4.42 & 2 & 313, 310\\
147584 & 6090 & $-0.06$ & 4.45 & 6 & 332\\
150248 & 5687 & $-0.11$ & 4.30 & 1 & 486, 134\\
156274 & 5242 & $-0.37$ & 4.40 & 1 & 163\\
157089 & 5785 & $-0.47$ & 4.09 & 1 & 182\\
159656 & 5845 & 0.09 & 4.32 & 2 & 357\\
160691 & 5695 & 0.23 & 4.02 & 1 & 263\\
162396 & 6026 & $-0.37$ & 4.08 & 1 & 39 \\
164595 & 5790 & $-0.04$ & 4.44 & 2 & 147\\
172051 & 5502 & $-0.16$ & 4.43 & 5 & 496\\
182572 & 5569 & 0.40 & 4.10 & 4 & 449\\
187237 & 5850 & 0.16 & 4.48 & 2 & 284\\
189567 & 5656 & $-0.26$ & 4.20 & 1 & 431\\
190248 & 5691 & 0.39 & 4.26 & 1 & 273\\
193307 & 6018 & $-0.34$ & 4.18 & 1 & 341\\
196378 & 5996 & $-0.44$ & 3.92 & 1 & 409\\
196755 & 5639 & 0.04 & 3.70 & 1 & 243\\
199288 & 5724 & $-0.60$ & 4.55 & 1 & 227\\
199960 & 5940 & 0.27 & 4.26 & 7 & 326\\
203608 & 6022 & $-0.67$ & 4.31 & 1 & 214\\
205420 & 6255 & 0.00 & 3.89 & 1 & 450, 171\\
206395 & 6305 & 0.23 & 4.38 & 1 & 256, 269\\
206860 & 6106 & $-0.04$ & 4.68 & 4 & 247\\
207043 & 5775 & 0.07 & 4.55 & 2 & 276\\
210918 & 5721 & $-0.09$ & 4.27 & 1 & 161\\
211415 & 5753 & $-0.25$ & 4.27 & 5 & 202\\
212330 & 5670 & $-0.02$ & 3.91 & 1 & 279, 232\\
215648 & 6178 & $-0.027$ & 3.97 & 1 & 451, 301\\
216436 & 5755 & 0.04 & 3.94 & 2 & 70\\
221287 & 6241 & $-0.02$ & 4.37 & 1 &236\\ 
221343 & 5755 & 0.04 & 4.05 & 2 & 177\\
222368 & 6200 & $-0.02$ & 4.13 & 1 & 810\\
BD+15 3364 & 5785 & 0.07 & 4.44 & 2 & 175 \\
\hline
\end{tabular}
\end{table}   

\begin{figure*}[h]
    \centering
    \includegraphics[width=16cm]{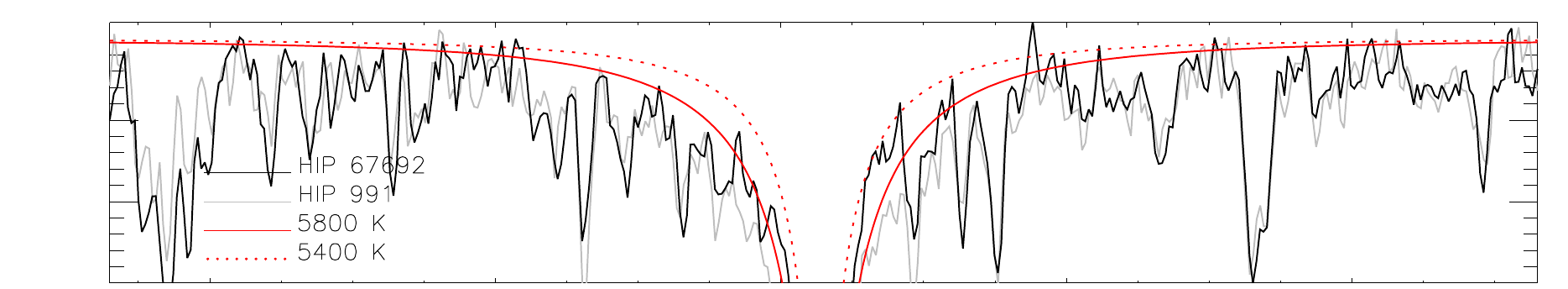}
    \includegraphics[width=16cm]{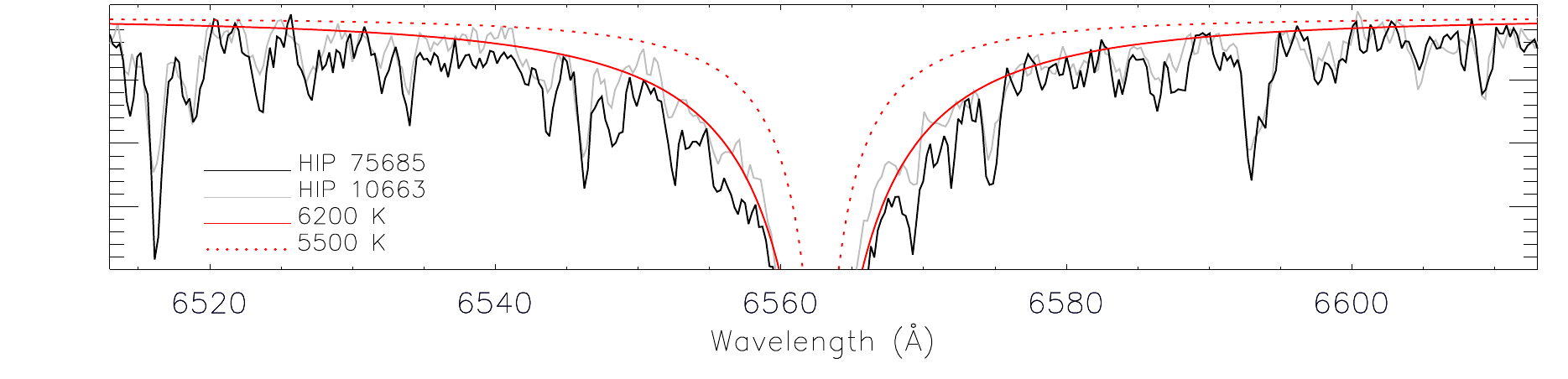}
    \caption{H$\alpha$ profiles of HIP 67692 (top) and HIP 75685 (bottom) compared to synthetic profiles with temperatures similar to their \teffPCA (full line)~and \teffP (dotted line). The observed profile of another candidate with \teffPCA~similar to, respectively, HIP 67692 and HIP 75685, is overplotted in gray.}
    \label{fig:Halpha}
\end{figure*}

\end{appendix}

\end{document}